\documentclass[12pt,preprint]{aastex}
\received{ }
\revised{ }
\accepted{ }
\usepackage{natbib}
\usepackage{color}

\shorttitle{QS \& CH in \ion{Mg}{2}~k line}
\shortauthors{Kayshap et al.}
\begin{document}
\title{Quiet-Sun and Coronal Hole in \ion{Mg}{2}~k line as observed by IRIS}

\author{Pradeep Kayshap}
\affil{Group of Astrophysics, University of Maria Curie-Sk{\l}odowska, ul. Radziszewskiego 10, 20-031 Lublin, Poland} 
\author{Durgesh Tripathi}
\affil{Inter-University Centre for Astrophysics, Post Bag - 4, Ganeshkhind, Pune-411007, India} 
\email{durgesh@iucaa.in}
\author{Sami K. Solanki}
\affil{Max-Planck Institute for Solar System Research, Justus-von-Liebig-Weg 3, 37077, G\"ottingen, Germany} 
\affil{School of Space Research, Kyung Hee University, Yongin, Gyeonggi-Do,446-701, Republic of Korea}
\author{Hardi Peter}
\affil{Max-Planck Institute for Solar System Research, Justus-von-Liebig-Weg 3, 37077, G\"ottingen, Germany} 

\begin{abstract}Coronal holes (CHs) regions are dark in comparison to the quiet-Sun (QS) at the coronal temperatures. However, at chromospheric and transition region (TR) temperatures, QS and CHs are hardly distinguishable. In this study we have used the \ion{Mg}{2}~2796.35~{\AA} spectral line recorded by the Interface Region Imaging Spectrometer (IRIS) to understand the similarities and differences in the QS and CH at chromospheric levels. Our analysis reveals that the emission from \ion{Mg}{2}~k3 \& k2v that originates in the chromosphere is significantly lower in CH than in QS for the regions with similar magnetic field strength. The wing emissions of \ion{Mg}{2}~k that originates from the photospheric layer, however, do not show any difference between QS and CH. The difference in \ion{Mg}{2}~k3 intensities between QS and CH increases with increasing magnetic field strength.  We further studied the effects of spectral resolution on these differences and found that the difference in the intensities decreases with decreasing spectral resolution. For a resolution of 11~{\AA}, the difference completely disappears. These findings are not only important for mass and energy supply from the chromosphere to the corona but also provides essential ingredients for the modelling of the solar spectral irradiance for the understanding of the Sun-climate relationships.
\end{abstract}

\section{Introduction}
  
Dark regions on the Sun at coronal temperatures are termed coronal holes (CHs), which can be observed both at low heliographic latitudes (primarily around solar activity maximum) and at the solar polar caps (primarily around solar activity minimum). At these temperatures, the emission from CHs is significantly lower than from other regions such as Quiet Sun (QS) and active regions \citep[see, e.g.,][]{Wal1975}, making them distinguishable from other areas. The difference between the two types of structures is thought to be a product of the different magnetic topology, with field lines being mainly closed, loop-like structures in the quiet Sun, while they possess an open funnel-like configuration in CHs. However, QS and CHs are not clearly distinguishable in intensity images taken at chromospheric temperatures except for \ion{He}{1} \& \ion{He}{2} lines, whose formation is very sensitive to EUV radiation from the corona, as well as to energetic electrons coming down from the corona, e.g., in the form of heat conduction \citep[see, e.g.,][]{Avrett1994, Andretta1997, Centeno2008, Leenaarts2016}. Therefore, the differences between CHs and QS seen in \ion{He}{1} \& \ion{He}{2} lines are likely linked to the differences in the overlying TR and corona. Based on the  Solar Ultraviolet Measurements of Emitted Radiation \citep[SUMER,][]{WilCM_1995} observations on board the Solar and Heliospheric Observatory (SoHO), \cite{WilDMH_2000} reported that the radiance of \ion{He}{1}~584~{\AA} line is about 30\% lower in CH than in QS. The differences have been noted also in other parameters such as Doppler shift and line widths \cite[see, e.g.,][]{DupPJ_1996,Peter_1999,Stucki2000, Stucki2002}.  

The H and K lines of \ion{Ca}{2}, that represent the chromospheric emission, are found to have the central intensities slightly enhanced in CH over QS \citep[see e.g.,][]{TepOT2006, Tep2007}. In another study, using multiple spectral lines from SUMER, \cite{Stucki2000} suggested that for cooler chromospheric lines like \ion{Ni}{2}, the spread of intensities in CHs is larger than in the QS. However, this difference disappeared in transition region lines, e.g., \ion{O}{4}. This behavior was reversed in coronal lines. Using the observations from the Coronal Diagnostics Spectrometer \citep{HarSC_1995} on board SoHO, it was found that the CHs and QS can be clearly distinguished only at temperatures higher than 7$\times$10$^5$~K, ignoring the He lines \citep{Stucki2002}. The weakness of such studies, particularly at chromospheric and TR temperatures is that they do not consider the strong dependence of radiance on magnetic flux. Thus, a difference in magnetic flux (or a distribution of the flux) in the CH relative to the QS could explain any differences found in the lower atmosphere. If, after taking any difference in magnetic flux into account there still is a difference between the intensity of chromospheric lines in CHs and QS, it would suggest that the magnetic field plays an important role in heating the quiet chromosphere. 

The fact that CHs are clearly visible in coronal, but not in chromospheric, radiation was explained by \cite{Wie2004} by the different density of long and high coronal loops, but the similar number of short, low cool loops in QS and CHs. A clear difference in chromospheric thermal structure between QS and CHs, other than in He lines, could indicate that the number density of short, cool loops is indeed different in CHs and QS, or that the heating of such loops is less efficient in CHs. 
 
The Interface Region Imaging Spectrometer \citep[IRIS,][]{DePon2014} provides high spectral, spatial and temporal resolution spectra in various spectral lines including the \ion{Mg}{2}~h \& k lines. Therefore, such data provide an excellent opportunity to compare, for the first time, the properties of these lines and their behavior in QS and CH regions. In particular, it is unknown if the \ion{Mg}{2}~h \& k lines have a different intensity in CH than in QS. If they do, it would shed some light on either their formation mechanism, or on the difference in properties of the the middle-upper chromosphere in CHs and QS. At present this layer is expected to be roughly similar in both types of magnetic topology. If a difference is found, then it would suggest that the chromosphere is at least partly magnetically heated and that the number density of small loops is different in the two regions. 

The rest of the paper is structured as follows. The details of the observation and data analysis are presented in \S\ref{section:obs}. In \S\ref{section:results}, we describe the observational findings of the present work. Discussion and conclusions are outlined in \S\ref{section:diss_conl}.

\section{Observation and Data-Analysis}\label{section:obs}

In the present study, we have used observations recorded with IRIS, and the Atmospheric Imaging Assembly \citep[AIA;][]{LemTA_2012} and the Helioseismic and Magnetic Imager \citep[HMI;][]{SchSB_2012} on board the Solar Dynamic Observatory (SDO). Since the primary aim of our study is to compare the radiance of the \ion{Mg}{2}~k line in QS and CH for magnetic elements with similar strength, it is important to consider the sets of observations where QS and CH are observed either in a single IRIS raster, or at least are observed at the same $\mu$ value (where $\mu$ is defined as $\mu$~=~cos$\theta$, with $\theta$ being the heliographic longitude) sufficiently close in time, so that there is no degradation of the instrument response between the two observations. This is necessitated to avoid any bias introduced by the strong centre to limb variation (CLV) in the radiance in \ion{Mg}{2} lines \cite[see e.g.,][]{Goutt1974,Avrett2013,Sch2015}. We have chosen sets of observations that fulfil these requirements for the study here. See Table~\ref{table_obs} for the details of observations. One of the observations recorded on November 10, 2015, contained a portion of QS as well as of a CH in a single raster. In the 2nd and 3rd sets of observations, the QS and CH were observed on two different dates, although at almost the same $\mu$ value. Here we concentrate on the detailed analysis of the first and 2nd set of observations. Since the analysis, as well as the results of the third set are very similar to the first two cases, we have relegated the results obtained using set 3 to an Appendix. 

\begin{figure*} 
\centering
\includegraphics[width=0.8\textwidth]{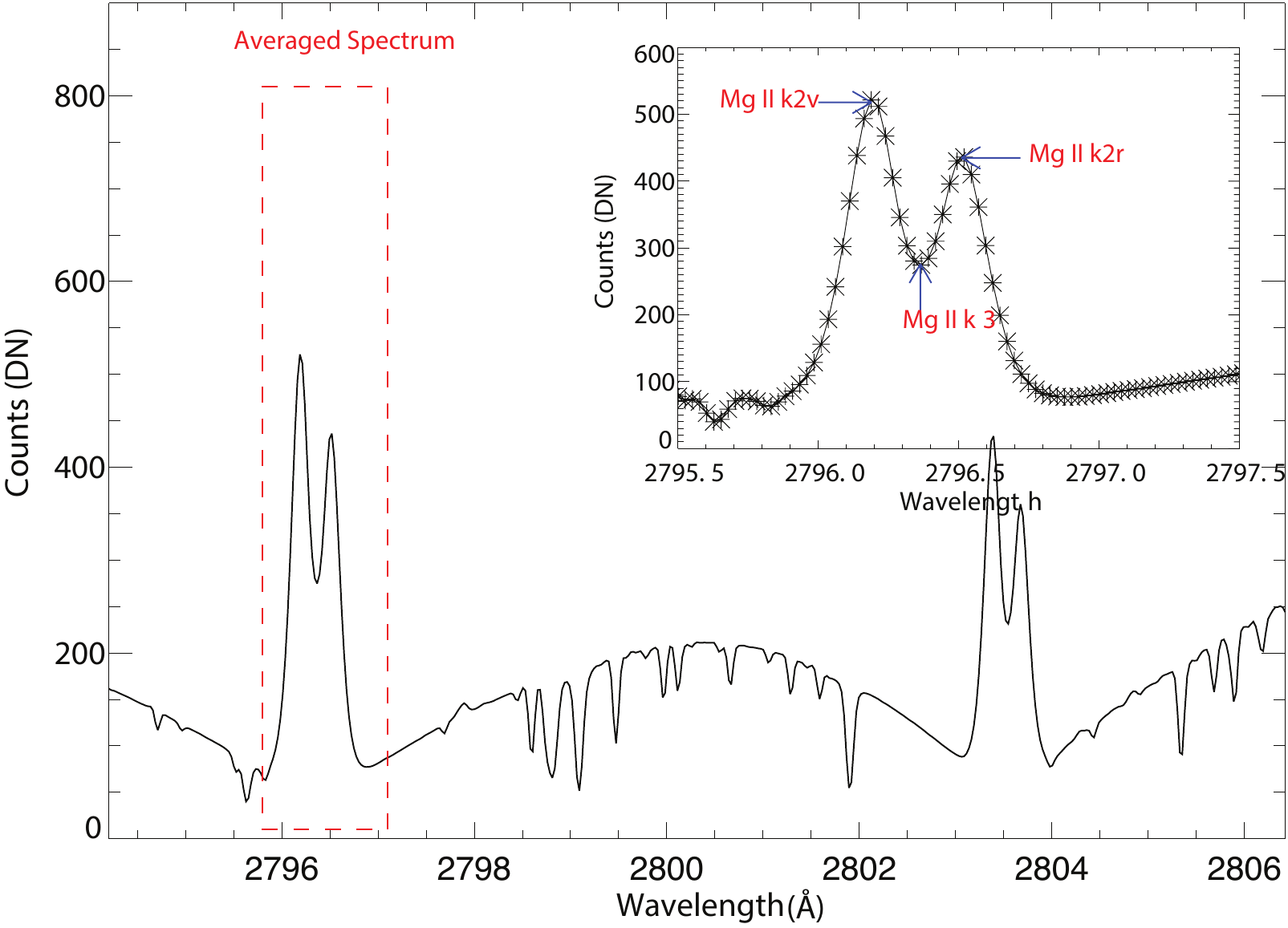}
\caption{An averaged, over entire raster, spectrum obtained in the \ion{Mg}{2} window of IRIS observation taken on Nov 14, 2015. The k-line is boxed in red and is also shown in the inset. The blue arrows in the inset locate the k2v and k2r peaks and the central absorption at k3.}
\label{fig:fig2}
\end{figure*}

The \ion{Mg}{2}~k~spectral line (2796.35~{\AA}) is an optically thick line, whose core is comprised of two peaks and a central reversal except at a few locations, e.g., in sunspot umbrae. Figure~\ref{fig:fig2} displays the spectrum in the \ion{Mg}{2} window averaged over the entire raster recorded on November 14, 2015. The red box locates the k-line of \ion{Mg}{2} that is also shown as an inset in Fig.~\ref{fig:fig2}. The two peaks, i.e., k2v (the short wavelength peak), and k2r (the long wavelength peak), and the k3 (dip in the line intensity) are labeled \citep[see][for more on nomenclature of the \ion{Mg}{2}~k lines]{Leen2013}.  

The intensities in the k2 and k3 features of \ion{Mg}{2} line are highly sensitive to magnetic field \citep{Leen2013}. Therefore, we performed a pixel-to-pixel correlation between the intensity maps of k2 and k3 and the projected LOS magnetic field and performed a comparison among intensities obtained for k2v and k3 locations of the \ion{Mg}{2} line in QS and CH for regions with similar magnetic field density. To find the \ion{Mg}{2} k2 and k3 intensities, we have used the procedure "iris$\_$get$\_$mg$\_$features.pro", which is  described in \cite{PerLD_2013} and \cite{LeePC_2013}.

To achieve the aims of the paper, we need to compare the radiance of the \ion{Mg}{2}~k line in QS and CH. To remove any bias due to different amounts of magnetic flux in the CHs and QS regions studied, we compare the dependence of \ion{Mg}{2}~k intensity on pixel averaged magnetic field strength. The latter value is taken from HMI magnetograms, which are recorded when the IRIS raster had reached its midpoint. This time will be referred to as the centre time. The HMI data was re-binned as per IRIS resolution to overcome different pixel sizes and spatial resolutions of IRIS and HMI. For such an analysis it is mandatory to have near perfect co-alignment between the magnetograms taken using HMI and the spectral images in \ion{Mg}{2} line core obtained by IRIS. We have used cross correlation to align IRIS and HMI images.

\section{Observational Results}\label{section:results}
\subsection{Data Set 1: Covering quiet Sun and coronal hole in the same raster map}\label{section:first_case}

\begin{figure} 
\centering
\includegraphics[width=1.0\textwidth]{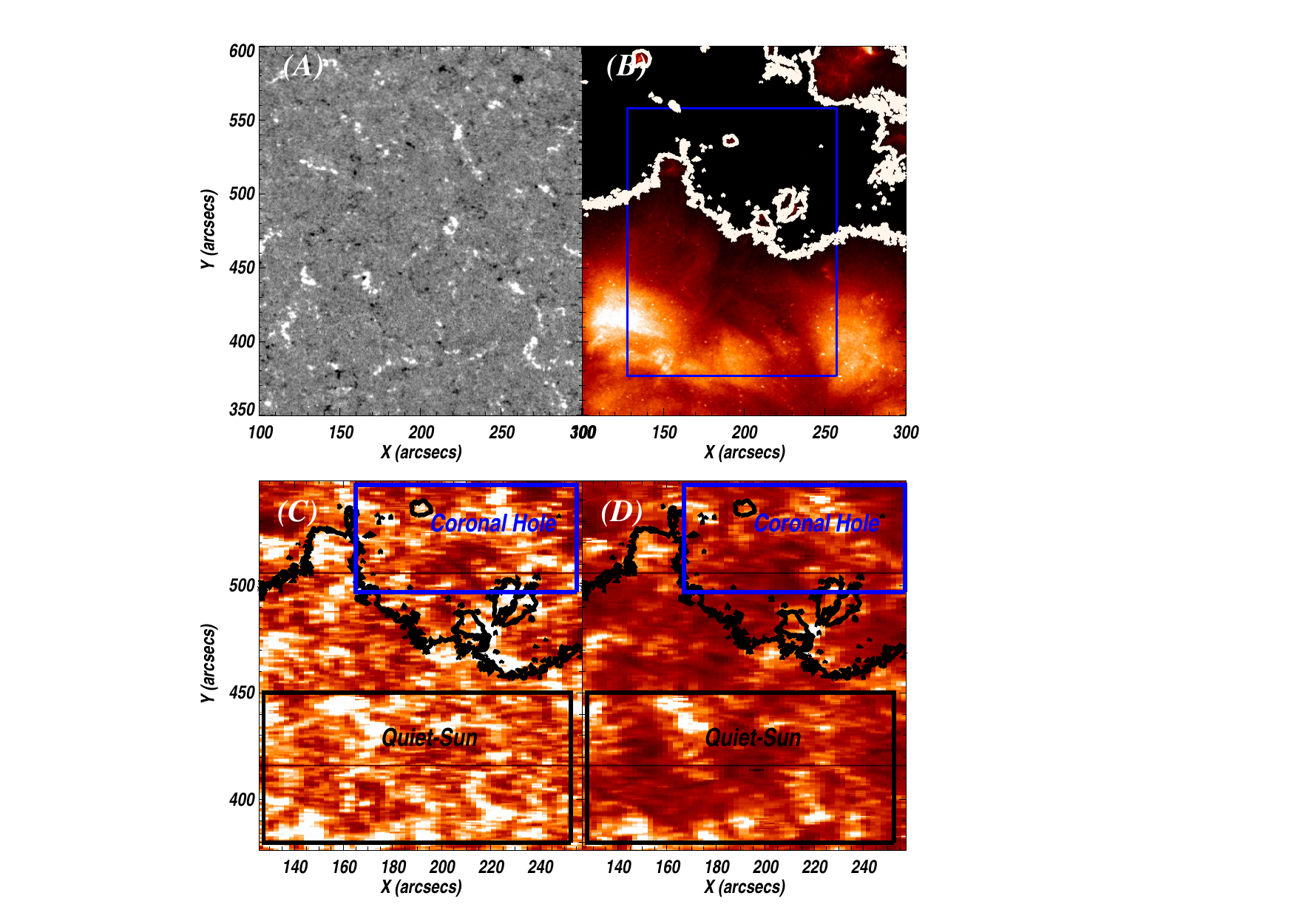}
\caption{Region of interest for data set 1. HMI magnetogram (panel A) and AIA~193~{\AA} image (panel B) showing a portion of the solar disk containing both QS and CH on 10 Nov 2015.  The blue box over-plotted on panel B represents the FOV of the IRIS raster. The IRIS intensity images obtained in \ion{Mg}{2}~k2v and \ion{Mg}{2}~k3 are shown in panels C \& D. The over-plotted contours (at a level of 80~DN/s in the AIA 193~{\AA} channel) in panels B, C \& D demarcate the boundary of the coronal hole and quiet Sun. The lower black and upper blue boxes in panels C \& D represent the QS and CH regions, respectively, which are used for further analysis. }\label{fig:fig1}
\end{figure}

\begin{figure} 
\centering
\includegraphics[width=1.0\textwidth]{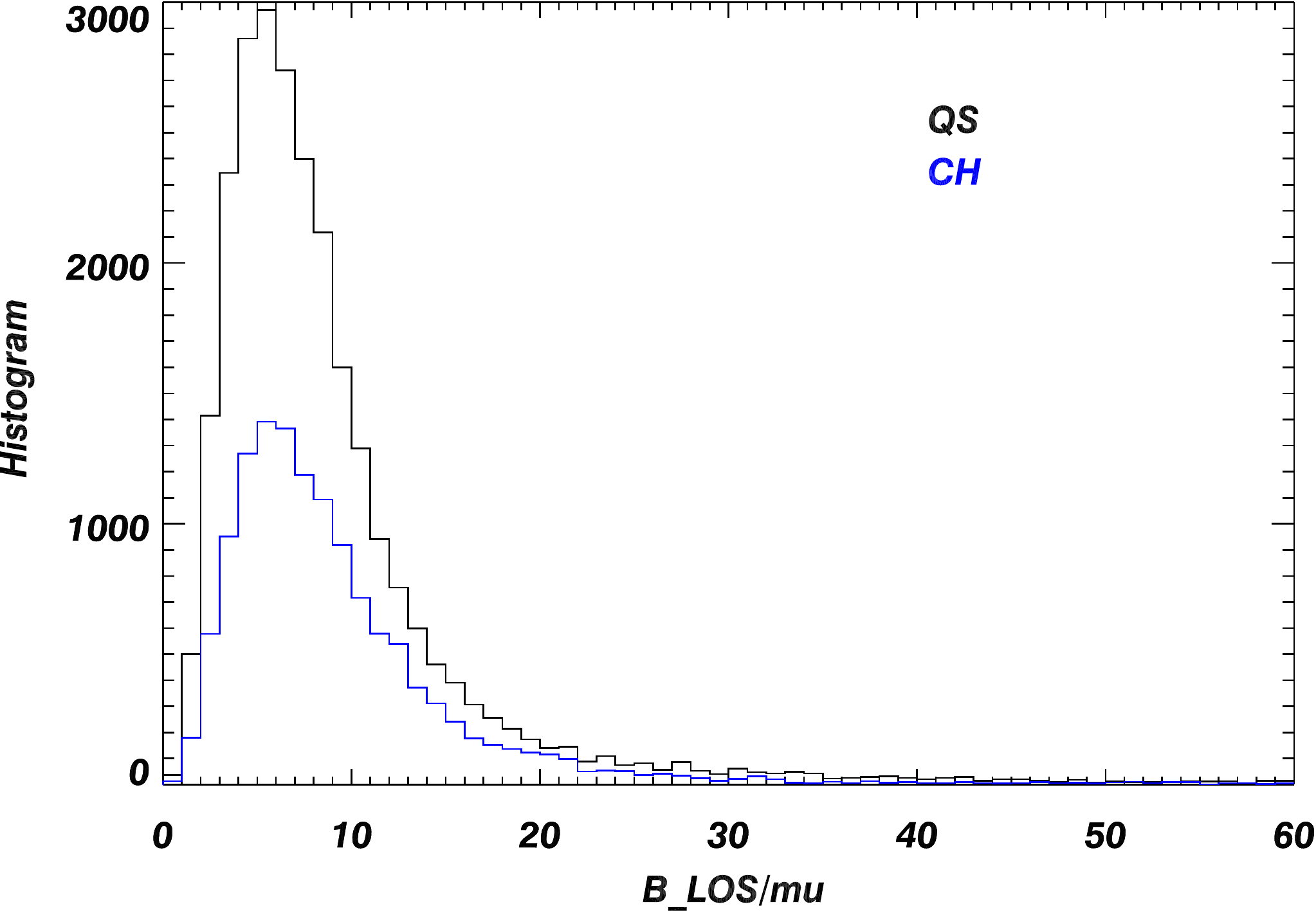}
\caption{Distribution of magnetic field in QS and CH as identified in Fig.~\ref{fig:fig1}.}\label{fig:mag}
\end{figure}

On November 10, 2015, IRIS rastered a region that covered a portion of QS as well as a CH using a 64 step coarse raster (i.e., 2{\arcsec} raster steps). Figure~\ref{fig:fig1} displays the HMI magnetogram (panel A) recorded at the center time of the IRIS raster and a portion of the Sun's disk imaged by AIA in its 193~{\AA} channel (panel B) on the same day. It is well established that CHs and QS are easily distinguished in coronal emission. Therefore, we have used AIA~193~{\AA} images to define the boundary between the QS and CH. By manual inspection, we have found that a threshold of 80~DN/pix demarcates the boundary between QS and CH reasonably well, as outlined by white contours on the AIA~193~{\AA} image (cf., panel B; Fig.~\ref{fig:fig1}). The blue box in panel B represents the region that was rastered by IRIS. It is clearly seen that the IRIS raster covered both QS and CH. Panels C and D are the maps representing the intensity at the (local) maximum and minimum of the spectral radiance of the \ion{Mg}{2} line profile at the k2v peak and the k3 dip, respectively. Clearly, both in k2v and k3, the structures in the QS are very similar to those in the CH region, unlike in the 193~{\AA} image. The same applies to the magnetic field obtained from HMI observations (panel A) that shows similar structure in CH and QS. For further quantitative analysis and comparison, we have selected the upper blue and the lower black boxes (shown in panels C \& D) representing CH and QS, respectively. Fig.~\ref{fig:mag} shows the distribution of magnetic field in QS (black) and CH (blue).

\begin{figure*} 
\centering
\includegraphics[width=1.0\textwidth]{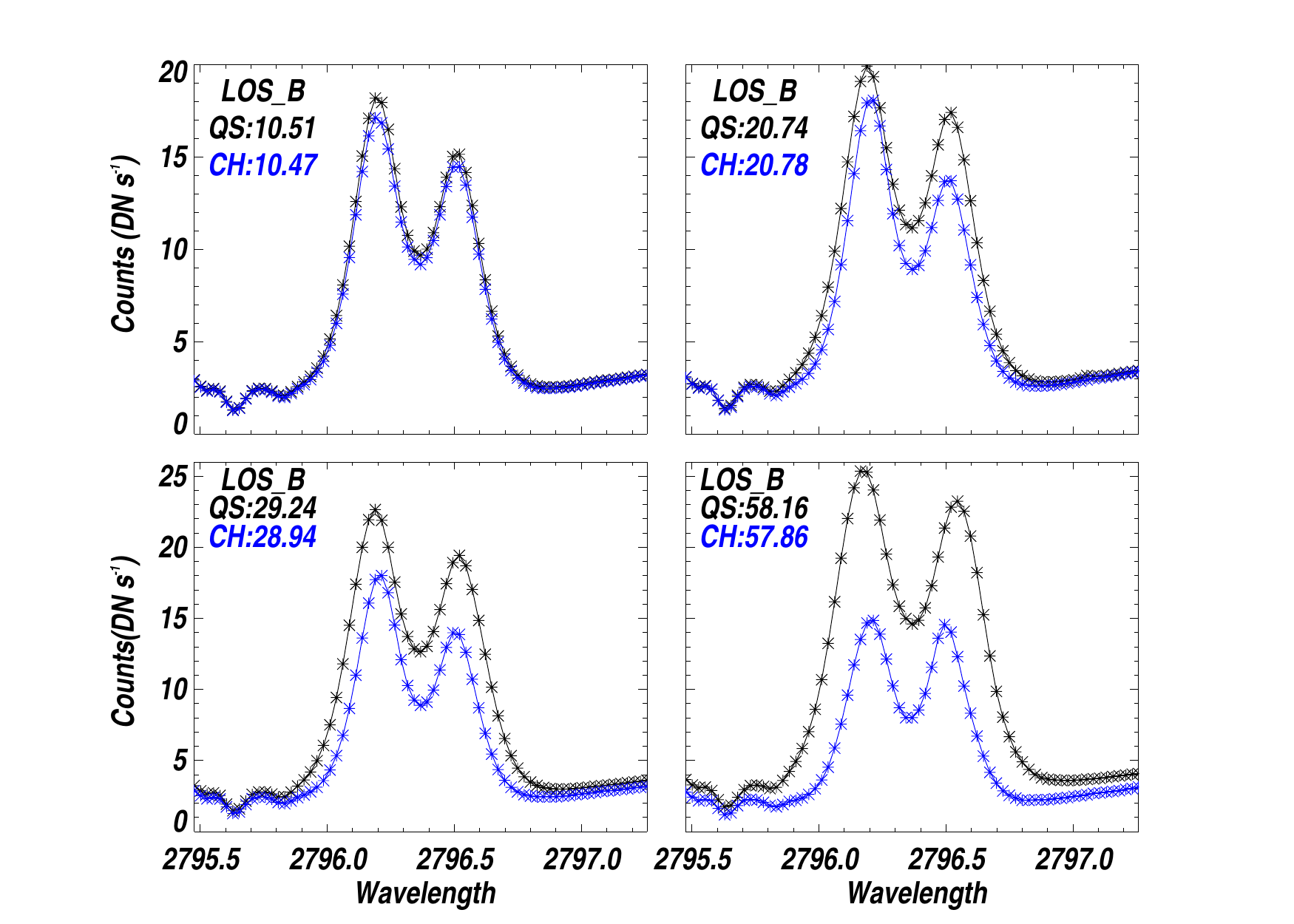}
\caption{\ion{Mg}{2}~k line profiles averaged over QS (black color) and CH (blue color) in four bins of different pixel averaged LOS magnetic field, as labeled in each panel. The values represents the average of $<$B$_{\rm LOS}$$>$/$\mu$ in the respective bin. The magnetic field units are in Gauss.}\label{fig:fig3}
\end{figure*}

To compare the strength of the \ion{Mg}{2}~k line in QS and CH for regions with very similar spatially averaged magnetic field strength, we grouped regions within a constant bin of 0.05 based on their pixel-averaged $<$B$_{\rm LOS}$$>$/$\mu$ on logarithmic  scale as obtained from the HMI magnetogram. A logarithmic scale was used  to have a sufficient number of pixels with high magnetic field. The number of \ion{Mg}{2} profiles in each magnetic field bin are given in Table~\ref{table_profile}. The complete QS and CH fields of view (FOV) marked by black and blue boxes in panels C \& D of Fig.~\ref{fig:fig1}, were divided into various bins covering pixel-averaged LOS magnetic field, $<B_{LOS}>$/$\mu$, from $\sim$2.0 Gauss up to $\sim$60.0 Gauss. We have excluded locations with higher magnetic field strength (i.e., more than 60 Gauss) as those are rare and introduce excessive scatter into the relationship at high field end. The spectra at all the locations in a given magnetic field bin are then averaged. Figure~\ref{fig:fig3} displays four averaged \ion{Mg}{2}~k line profiles for QS (blue) and CH (black) obtained in four bins of different $<B_{LOS}>$/$\mu$ as labeled in each panel. The plots clearly demonstrate that the \ion{Mg}{2} line is stronger in QS than in the CH, when compared at the same magnetic field strength. Moreover, the difference in the k2 peaks and the k3 dip increases with increasing magnetic field. 

\begin{figure*} 
\centering
\includegraphics[width=0.65\textwidth]{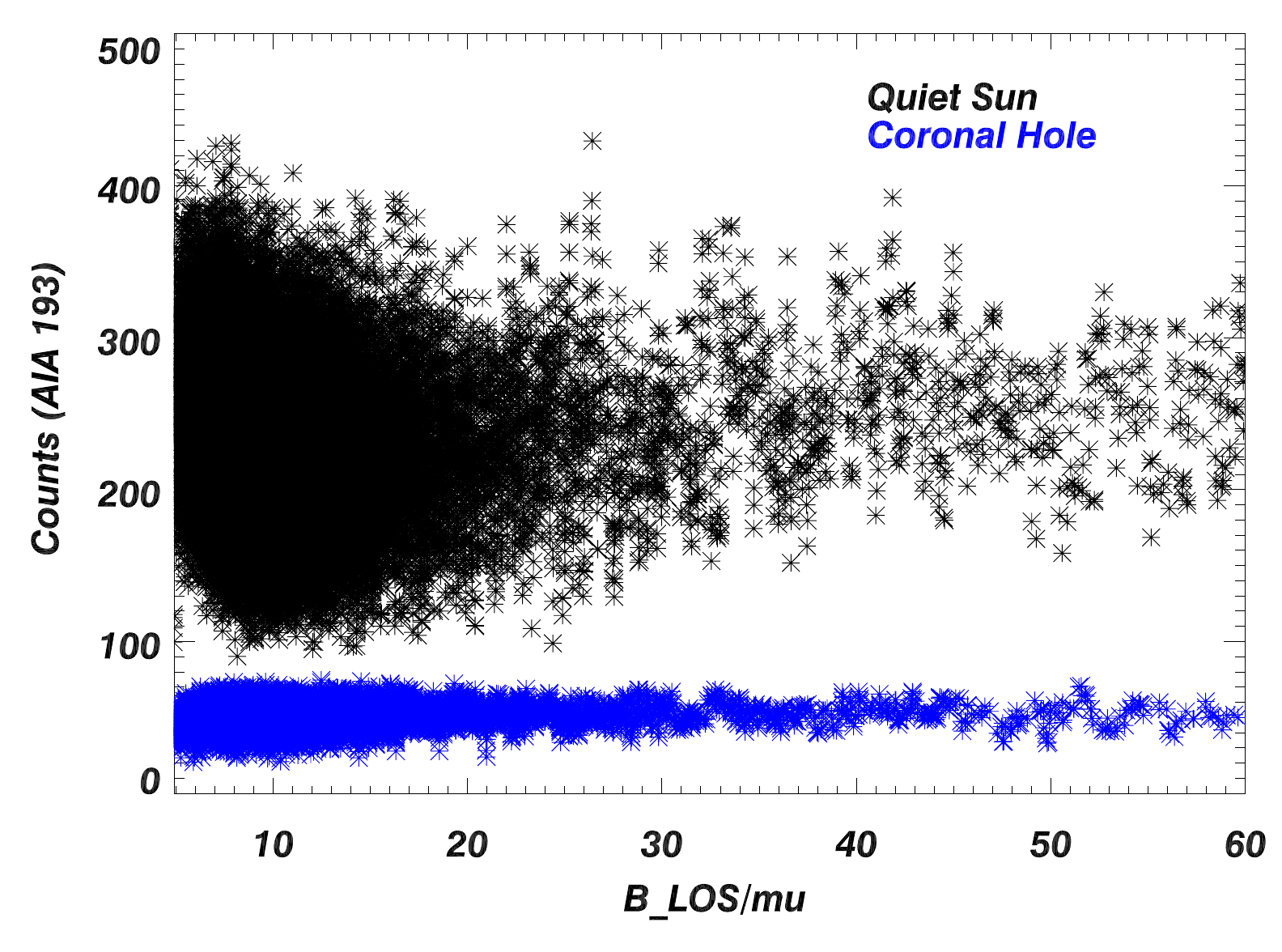}
\includegraphics[width=0.65\textwidth]{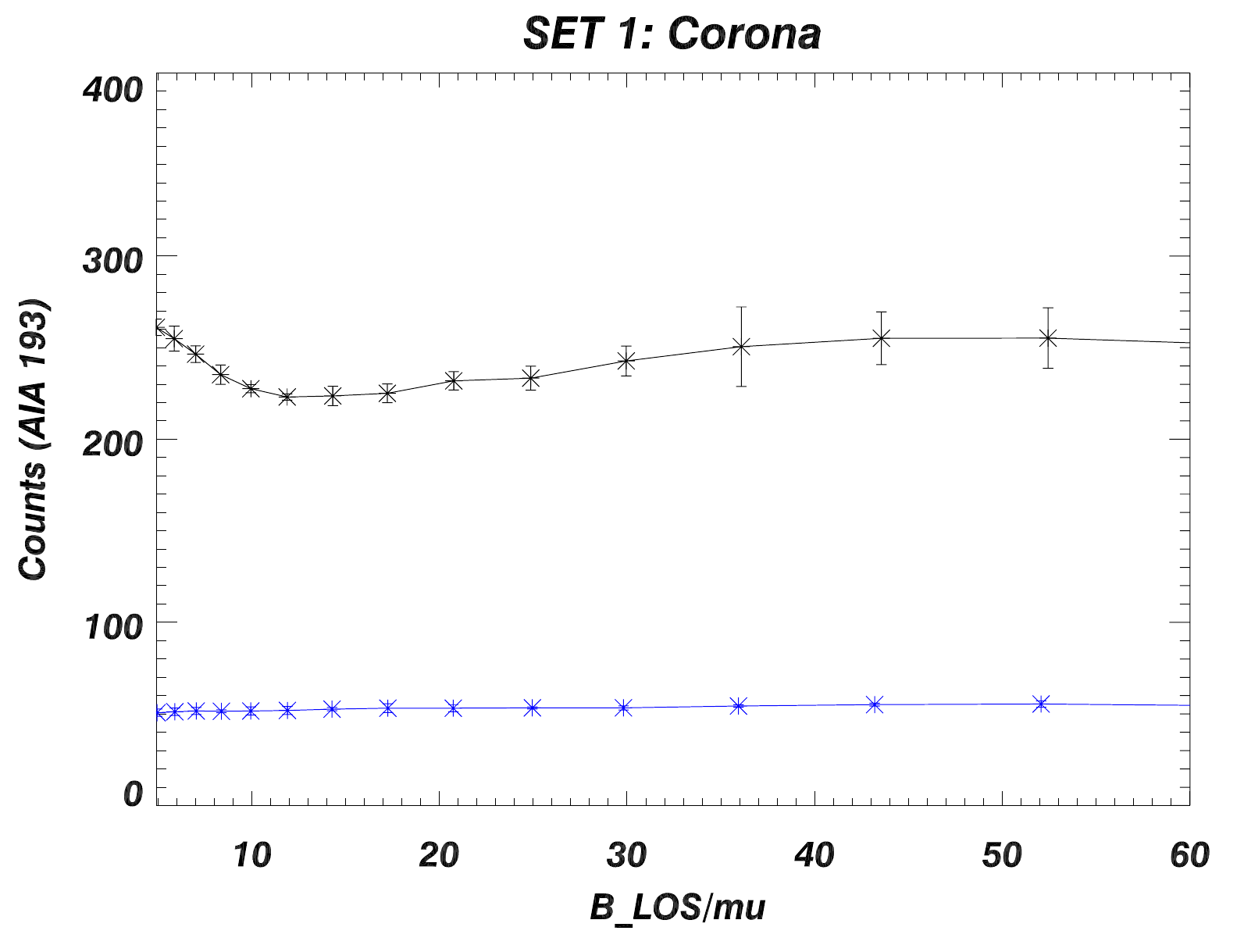}
\caption{Coronal intensities vs. magnetic field. Top panel: Intensity scatter plot as a function of magnetic field for quiet-Sun (black) and coronal hole (blue) as obtained using AIA~193{\AA}. Bottom panel: AIA~193{\AA} intensity averaged over bins of $<$B$_{\rm LOS}$$>$/$\mu$ in the quiet-Sun (black) and coronal hole (blue).} 
\label{fig:fig31}
\end{figure*}

We first investigate the relation of the coronal intensity to the magnetic field, which should show a clear difference between CH and QS, of course. The upper panel in Fig.~\ref{fig:fig31} displays the scatter plot of the intensities in QS (black) and CH (black) based on the AIA~193~{\AA} observations as a function of $<$B$_{\rm LOS}$$>$/$\mu$. The bottom panel shows the behavior of coronal intensities in QS (black) and CH (blue) as observed in 193~{\AA} after averaging the intensities in each bin of $<$B$_{\rm LOS}$$>$/$\mu$. As stated above we have selected a constant step of 0.08 $<$B$_{\rm LOS}$$>$/$\mu$ for each bin. These plots demonstrate the difference in QS and CH that is clearly discernible (see panel B in Fig.~\ref{fig:fig1}). The coronal QS intensities are a factor of more than four higher than those in the CH. The difference in the intensities first decreases and later increases with increasing $<$B$_{\rm LOS}$$>$/$\mu$, although this may be by chance (poor statistics), as we shall see by comparing with the other data sets analyzed here. 

\begin{figure*} 
\centering
\includegraphics[width=0.5\textwidth]{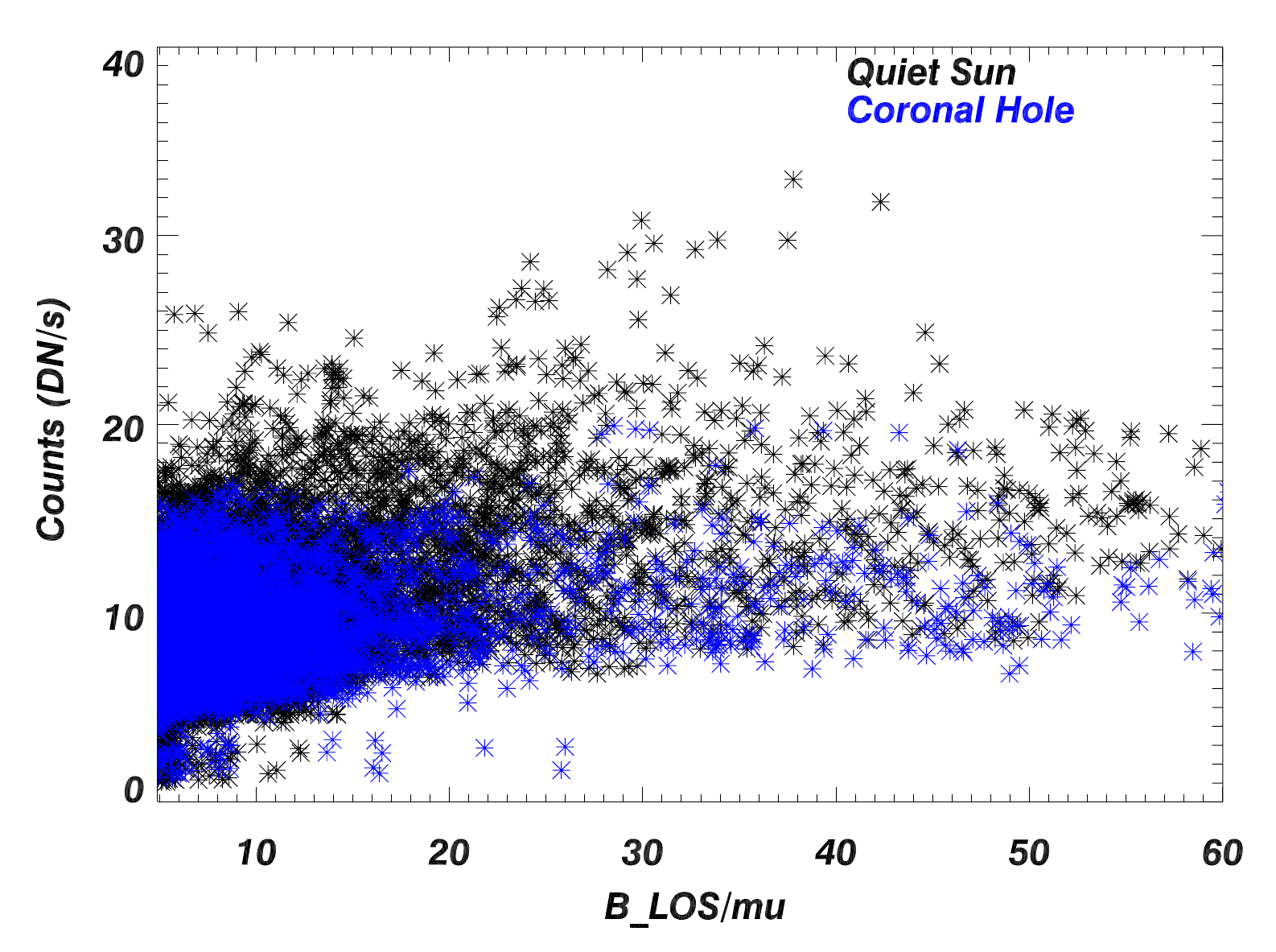}
\includegraphics[width=0.5\textwidth]{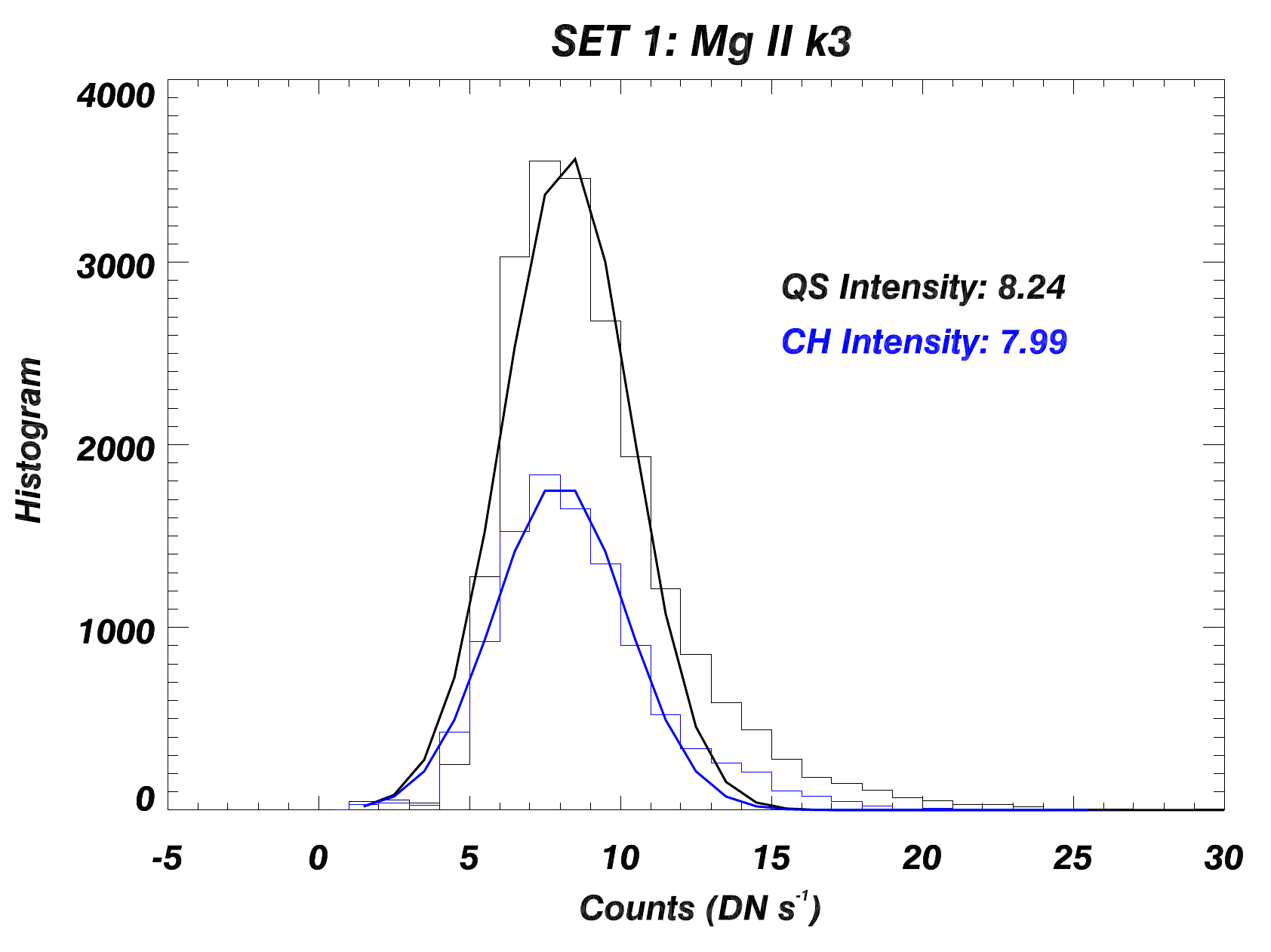}
\includegraphics[width=0.5\textwidth]{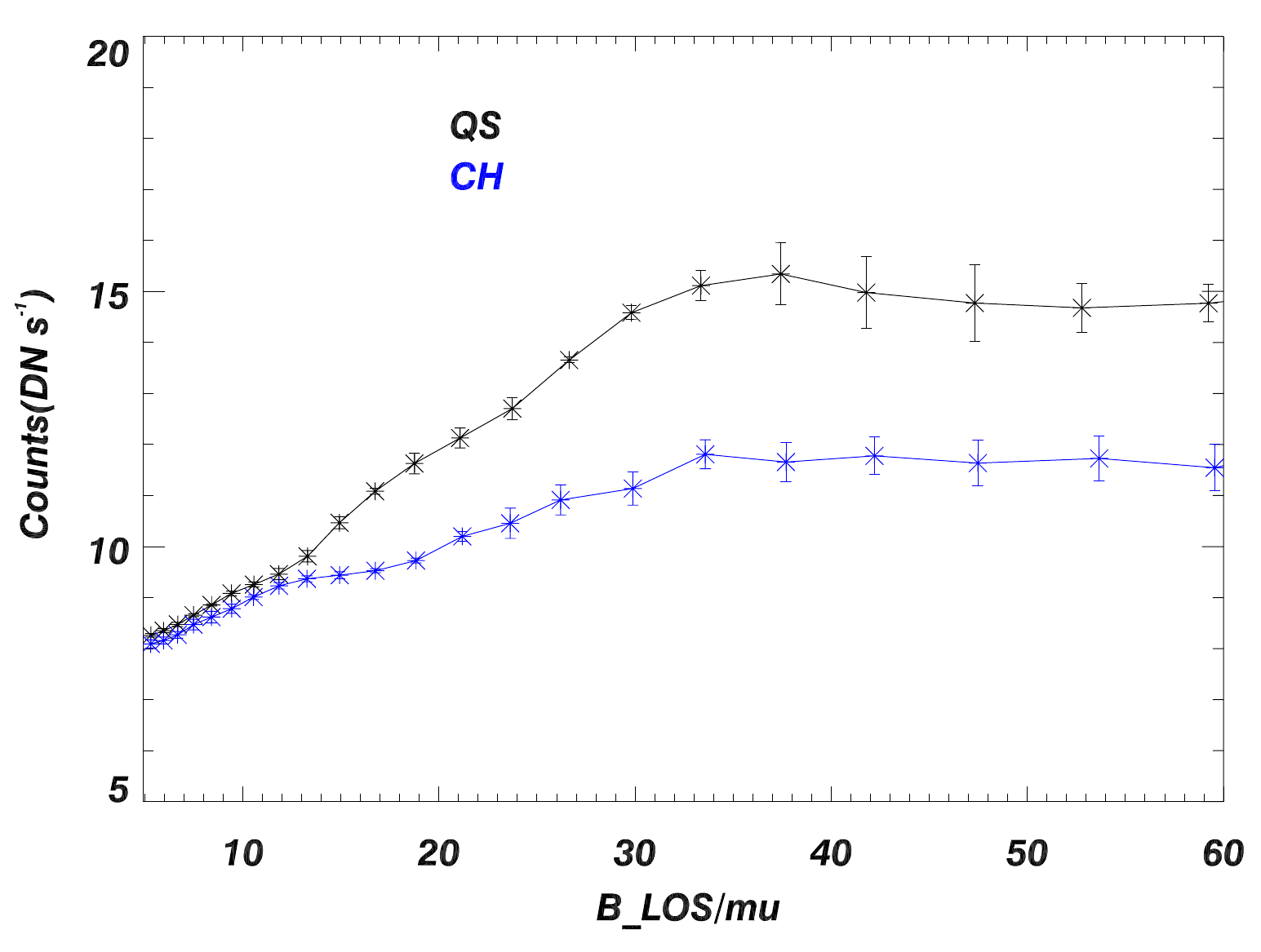}
\caption{Chromospheric intensity in \ion{Mg}{2} k3 vs. magnetic field. Scatter plot (top), histogram (middle) and averaged intensities with magnetic field (i.e., B/$\mu$) of QS (black) and CH (blue) for \ion{Mg}{2}~k3 (bottom).} 
\label{fig:fig32}
\end{figure*}

\begin{figure*} 
\centering
\includegraphics[width=0.5\textwidth]{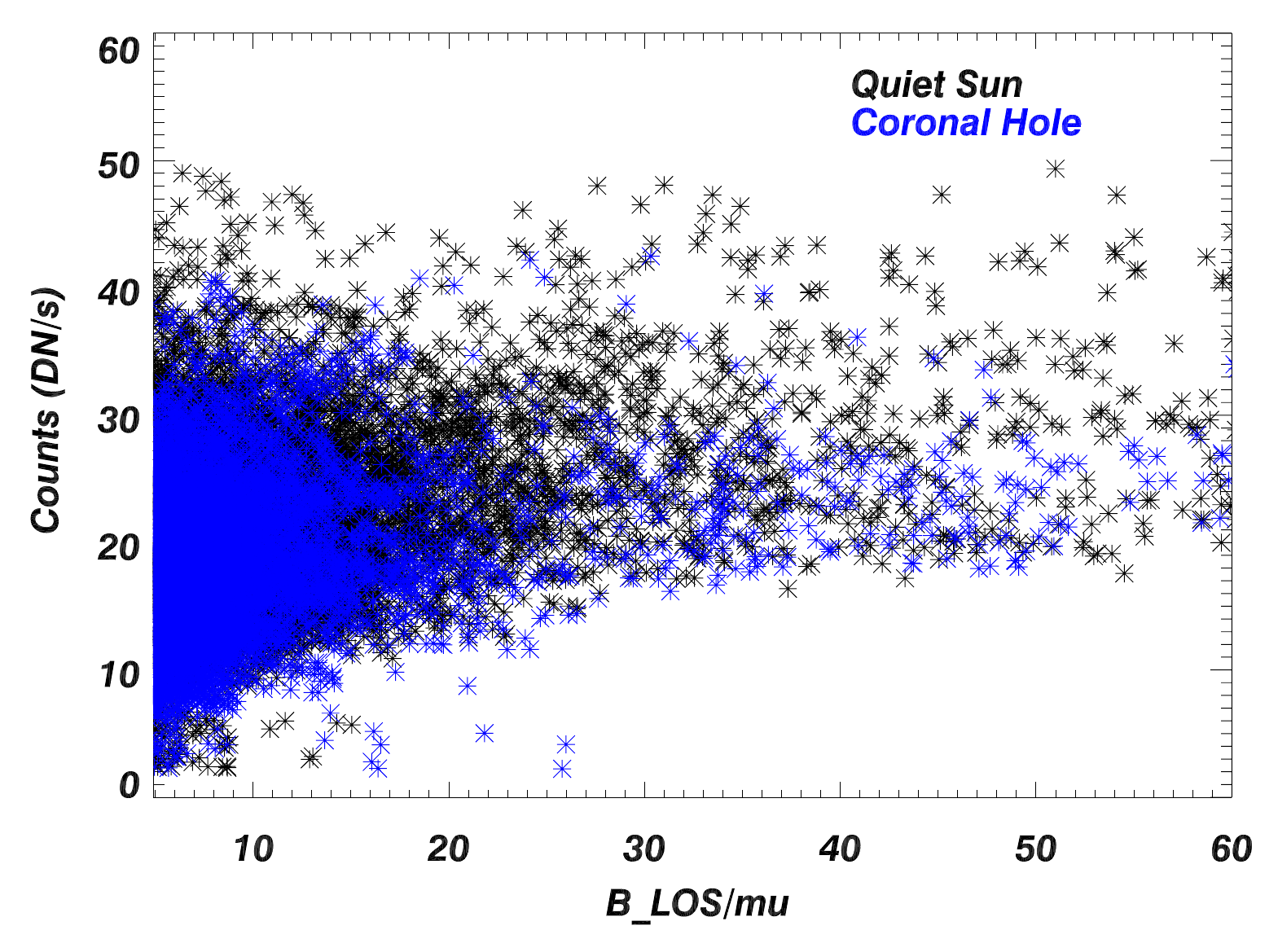}
\includegraphics[width=0.5\textwidth]{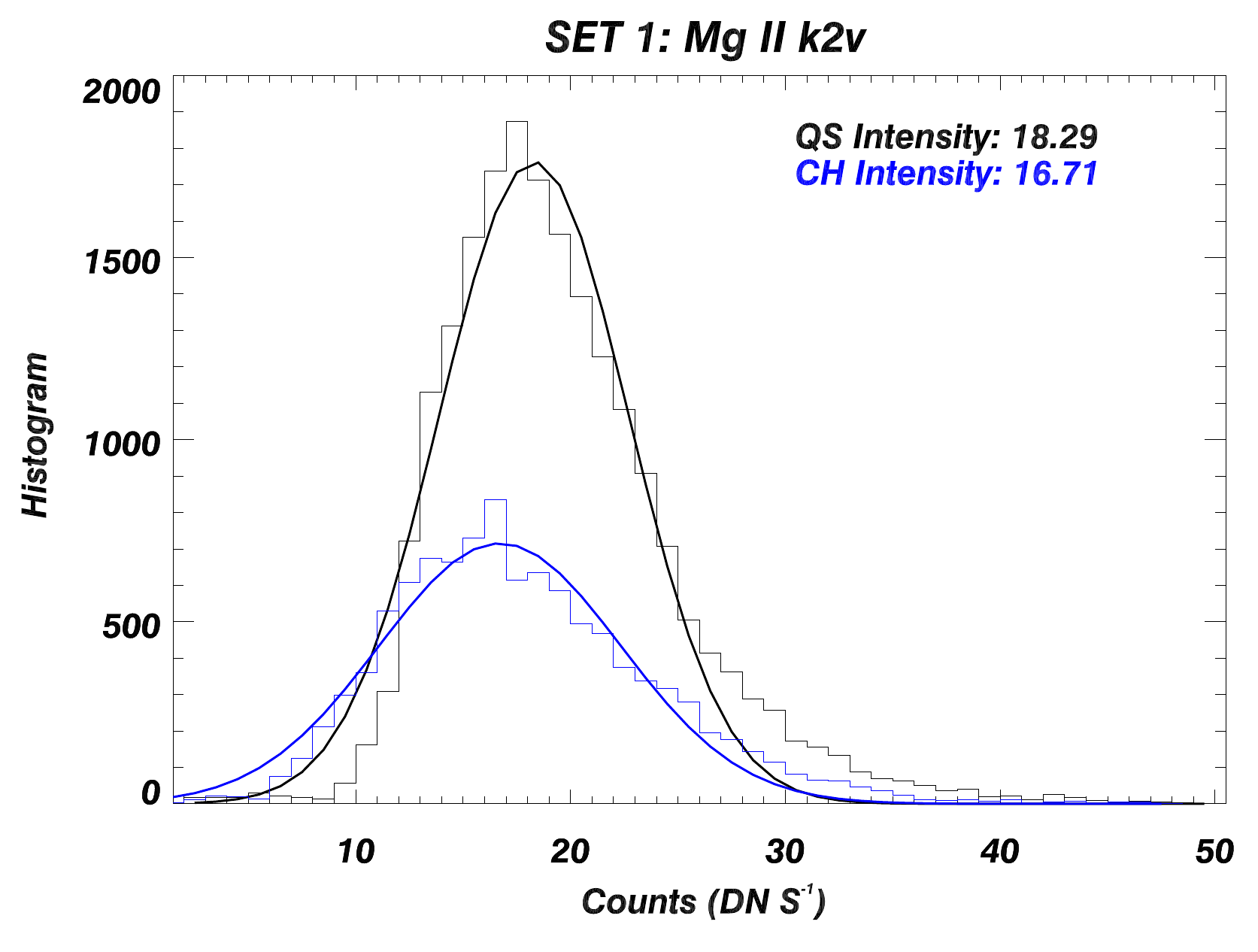}
\includegraphics[width=0.5\textwidth]{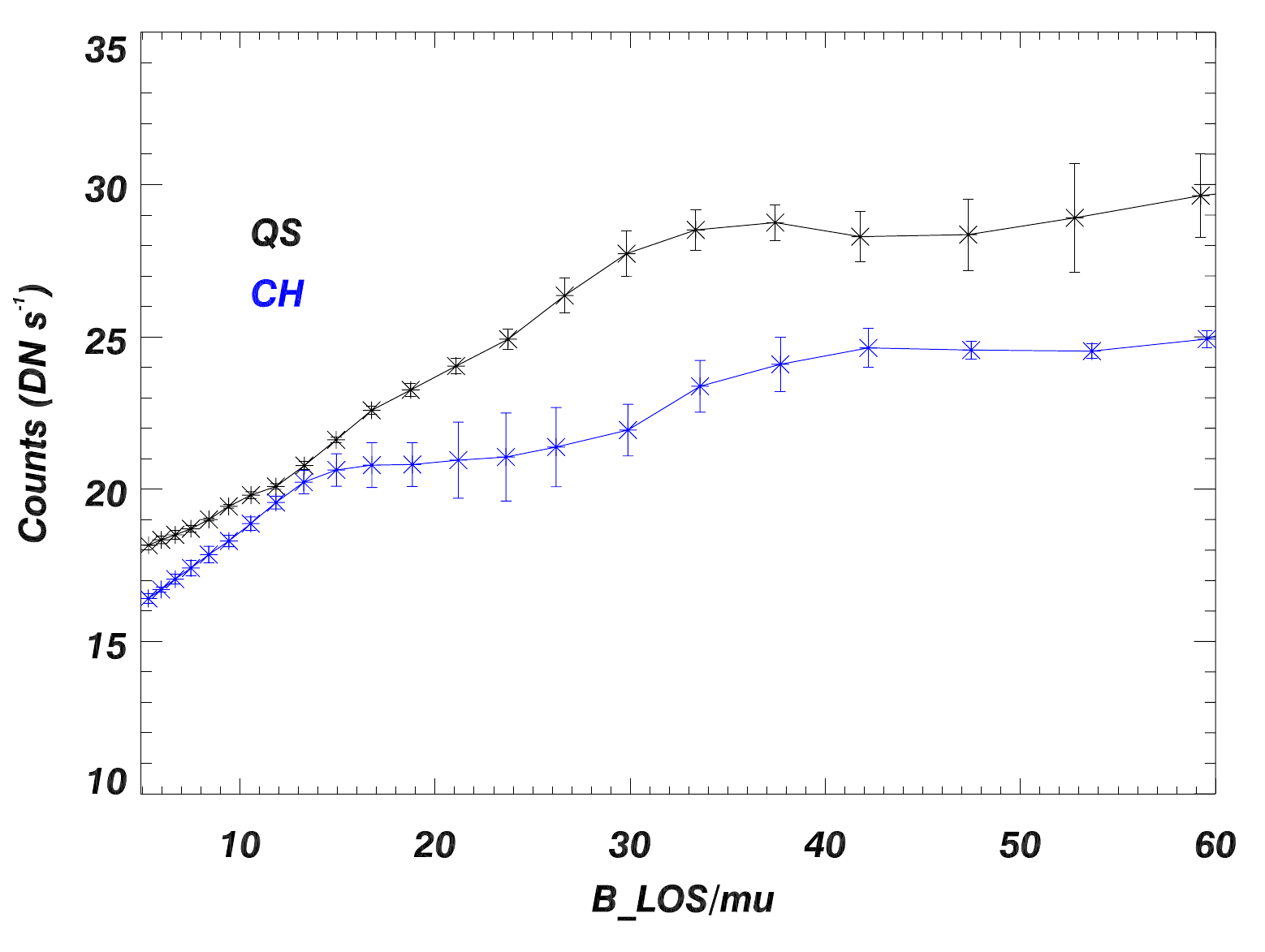}
\caption{Same as Fig.~\ref{fig:fig32} but for the \ion{Mg}{2}~k2v peak intensity vs. magnetic field.}
\label{fig:fig33}
\end{figure*}

The difference between CH and QS in \ion{Mg}{2} is best illustrated by relating the intensity of the \ion{Mg}{2} features to the magnetic field strength. Figures~\ref{fig:fig32} \&  \ref{fig:fig33} display scatter plots of intensities (top panel), and average intensities in different magnetic field bins (bottom panel) as a function of $<$B$_{\rm LOS}$$>$/$\mu$ of the QS (black) and CH (blue) regions shown in Figure~\ref{fig:fig1} for the k3 reversal and the k2v peak, respectively. The histograms of intensities are plotted in the middle panels. The scatter plots show that although most of the points overlap for similar magnetic field strength, there are regions in QS that are brighter than all regions in the studied CH. The histograms of QS and CH intensities in Figs.~\ref{fig:fig32} and \ref{fig:fig33} confirm that there is a large overlap (although it is smaller than the histograms suggest, as the dependence on $<$B$_{\rm LOS}$$>$/$\mu$ enhances it). The average QS intensities in the same magnetic field bins are larger than in CHs, and this difference increases with increasing magnetic field for both k3 as well as k2v. At even higher magnetic fields, a decrease in the difference is seen. We attribute this to poor statistics as there are very few locations with magnetic field higher than 50~Gauss. However, we cannot rule out that the stronger network elements display similar brightness in CHs and QS, while it is only the weaker elements that are different. We emphasize here that these differences are not visible when considering the images directly by eye.

\subsection{Data Set 2: quiet Sun and coronal hole observed in separate rasters}
\begin{figure*}  
\centering
\includegraphics[width=0.85\textwidth]{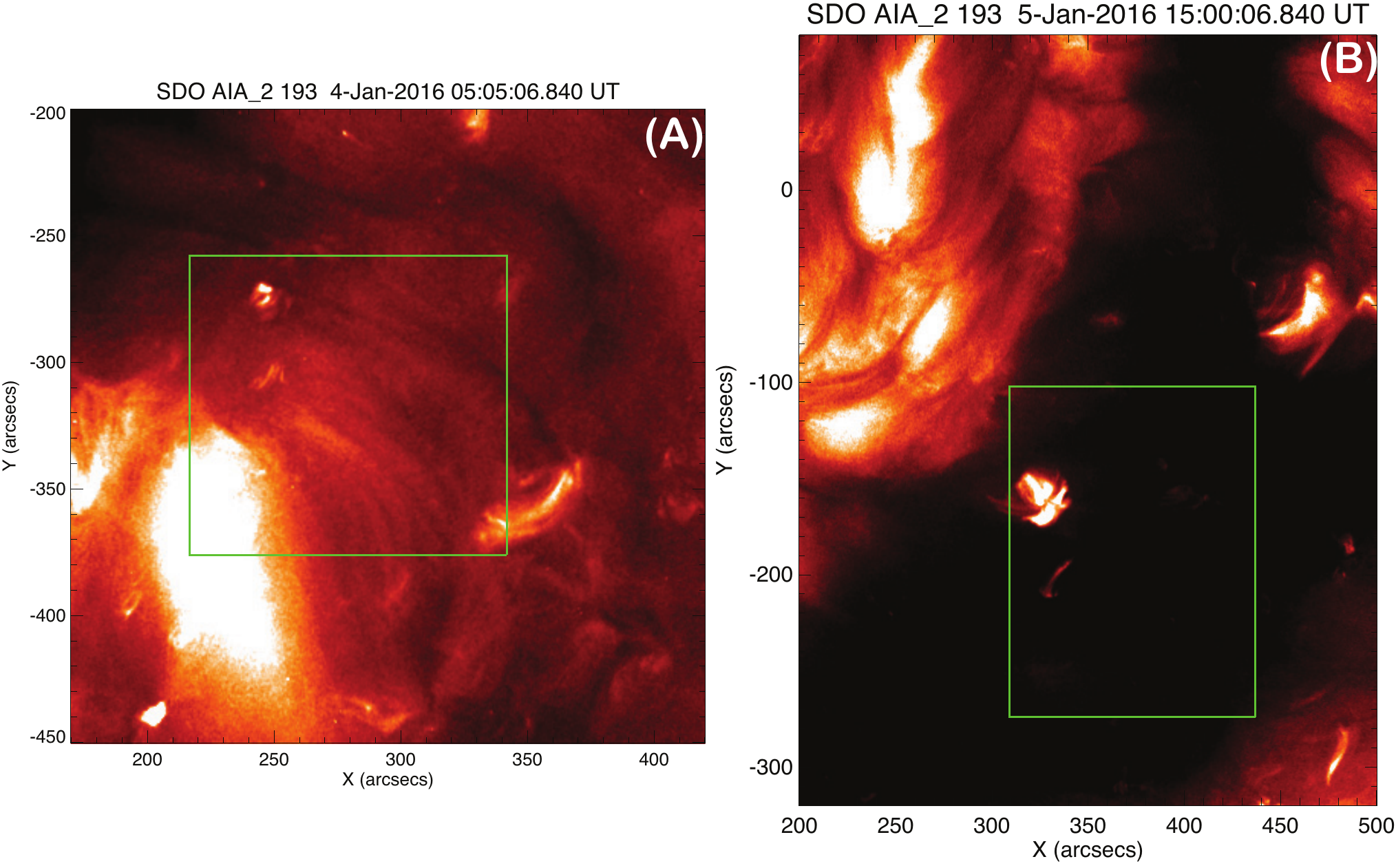}
\caption{Region of interest for data set 2. Images taken by AIA in the 193~{\AA} channel showing QS (left panel) and CH (right panel). The over-plotted green boxes locate the regions which were rastered by IRIS. Note that we have used the same minimum and maximum values to plot these images.}
\label{fig:fig4}
\end{figure*}

\begin{figure*}  
\centering
\includegraphics[width=0.65\textwidth]{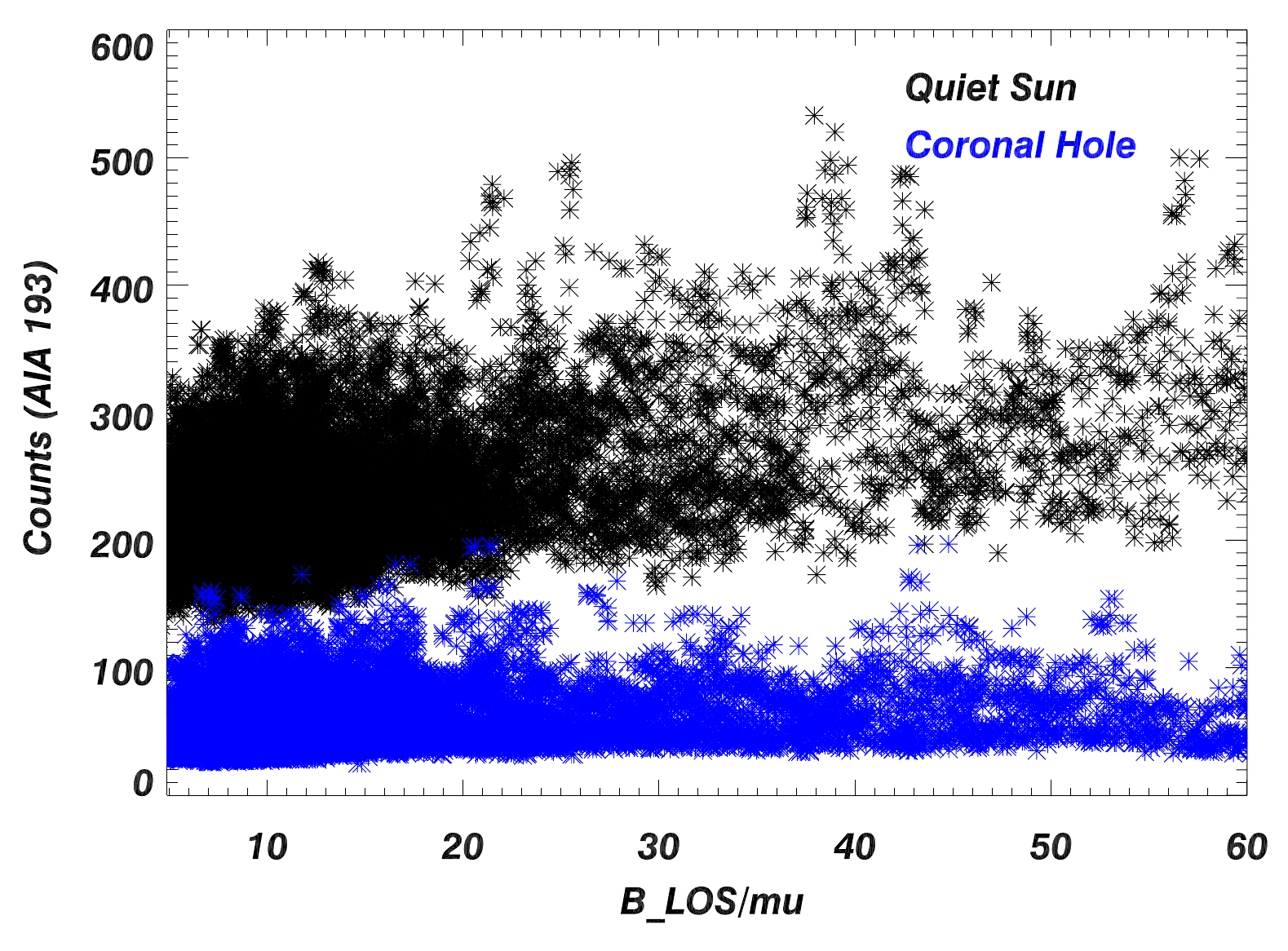}
\includegraphics[width=0.65\textwidth]{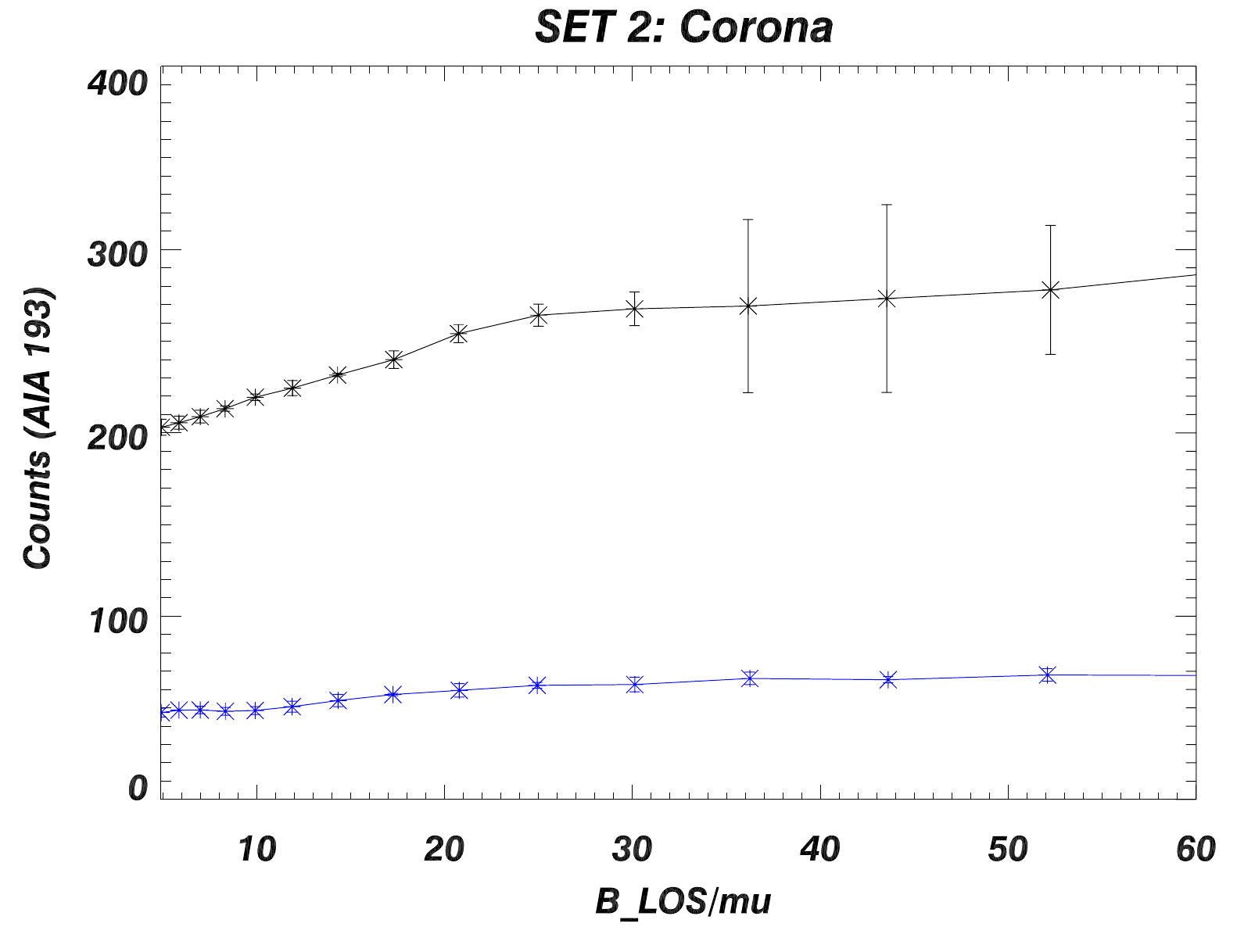}
\caption{Same as Fig~\ref{fig:fig31} but for case 2, showing coronal intensity vs. magnetic field.} \label{fig:fig5}
\end{figure*}

The second data set is comprised of two coarse rasters, one of QS taken on 2016 Jan 04, the other of a CH taken on 2016 Jan 05. The time difference between QS and CH observations is just $\sim$36 hours, which is small enough to leave our results unaffected by instrument degradation. Figure~\ref{fig:fig4} displays a portion of the Sun imaged by AIA~193~{\AA} on Jan 4th showing the observed QS region (panel A) and on Jan 5th showing CH (panel B). Over-plotted green boxes on both the images mark the areas covered by IRIS. The QS and CH regions mapped by IRIS were located at very similar heliographic coordinates ($\mu$ $=$ 0.9 both for QS and CH). The difference in the coronal intensities in QS and CH is evident from the images. Similar to the case 1 studied in \S\ref{section:first_case}, we produce a plot of intensities in QS (black) and CH (blue) using AIA 193~{\AA} observations as a function of $<$B$_{\rm LOS}$$>$/$\mu$. It is discernible that the QS region is brighter by roughly a factor of four for areas with a similar magnetic field, similar as observed in case 1 (see Fig.~\ref{fig:fig5}).

\begin{figure*} 
\centering
\includegraphics[trim = 7.2cm 0.1cm 7.2cm 2.0cm, scale=0.8]{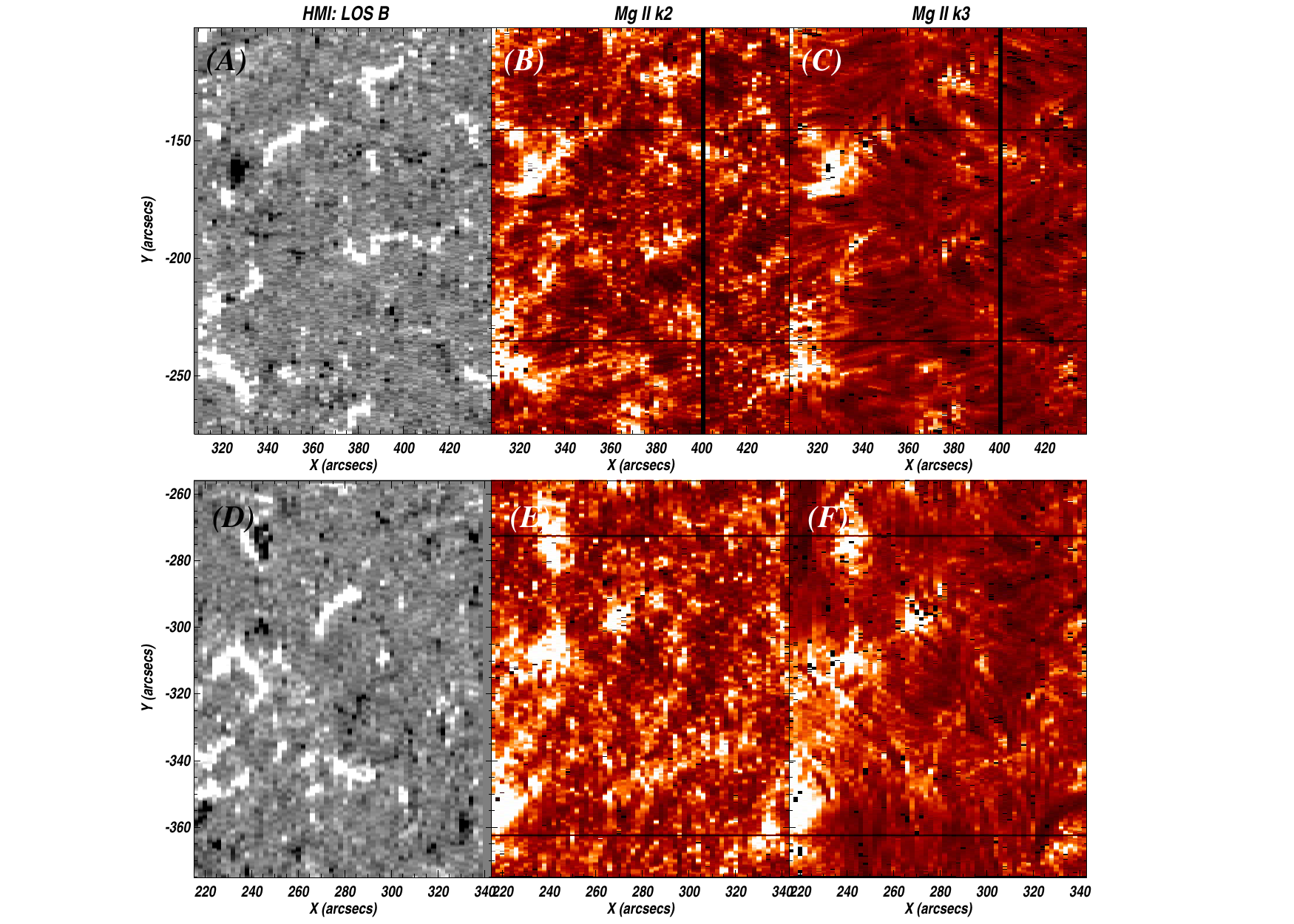}
\caption{Photospheric magnetic field and chromospheric emission for data set 2. Magnetograms are shown in panels A \& D, \ion{Mg}{2}~k2v maps are shown in panels B \& E and k3  maps is shown panels C \& F. Top row is for CH and bottom row is for QS.} \label{fig:fig6}
\end{figure*}

\begin{figure*}   
\centering
\includegraphics[trim=3.0cm 1.0cm 3.0cm 1.5cm,width=0.9\textwidth]{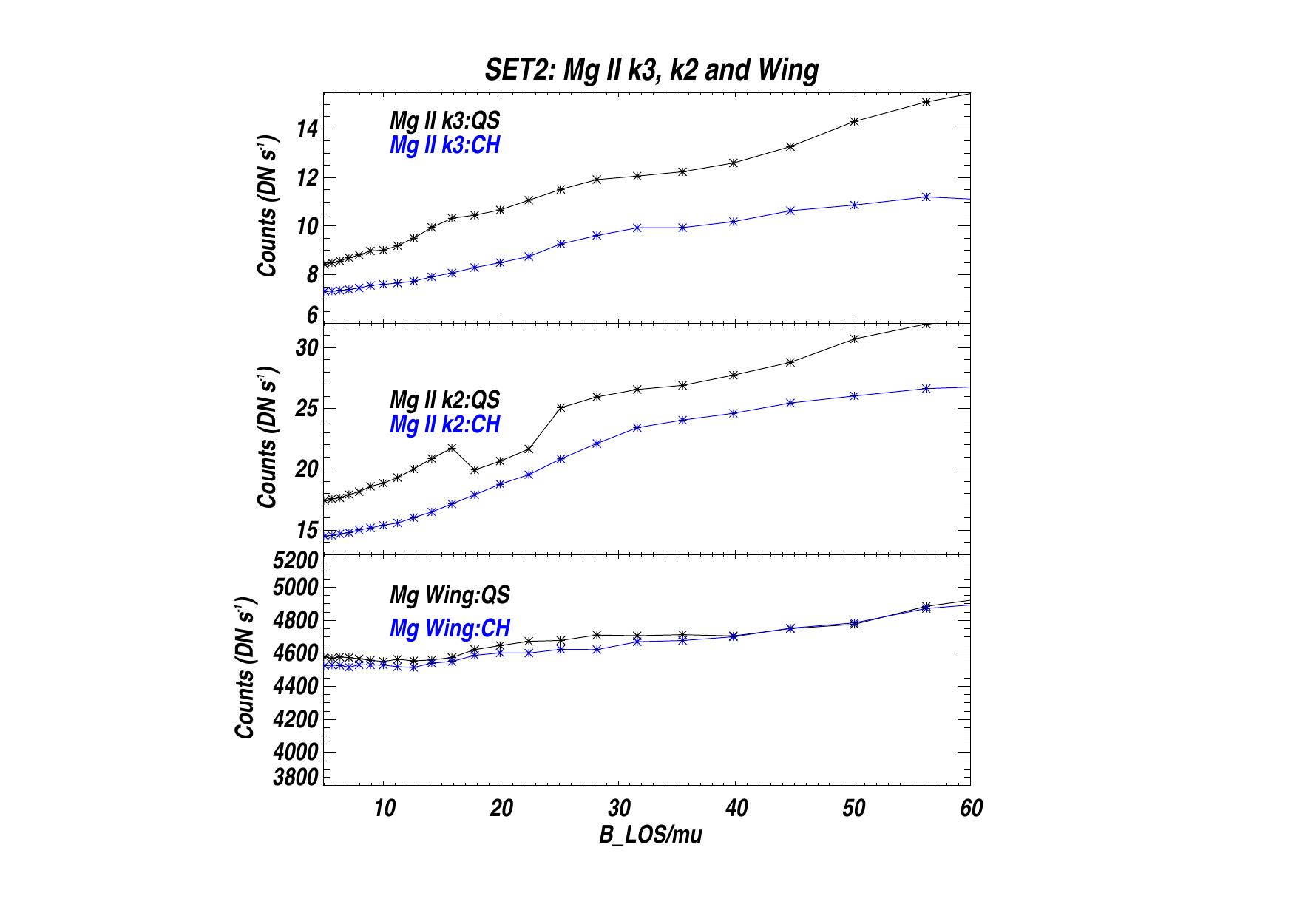}
\caption{Relation of chromospheric emission to magnetic field for data set 2. Average intensities obtained in \ion{Mg}{2}~k3 (top-panel), k2 (middle panel) and wing as a function of $<$B$_{\rm LOS}$$>$/$\mu$.} \label{fig:fig7}
\end{figure*}

Basically, the analysis of this case confirms the results for case 1 discussed earlier. Figure~\ref{fig:fig6} displays the magnetic field maps (A \& D) and intensity maps obtained for k2v (B \& E) and k3 (C \& F) for CH (top row) and QS (bottom row). As can be inferred from the intensity maps, there are no apparent visual differences in coronal holes and QS intensities in \ion{Mg}{2} as well as in general the magnetic field distribution. To perform a quantitative comparison, we produced similar plots for k3, k2v as we did for \textsl{Data~Set~1}. We additionally also consider the line wing covering the wavelenght range 2832.0-2834.0~{\AA} . The spectral window for \textsl{Data~Set~1} did not cover the \ion{Mg}{2}~k 2796.0~{\AA} line wing, therefore, we could not perform the wing analysis for $set 1$. Figure~\ref{fig:fig7} displays the averaged intensity in each bin of $<$B$_{\rm LOS}$$>$/$\mu$ for k3 (top), k2v (middle) and wing (right) of the \ion{Mg}{2} line. These plots reveal that the average QS intensities of k3 and k2v are larger than those for CH. The difference in the intensities increases with increasing magnetic field for k3, but that is not seen for k2v. 

In principle, the difference in \ion{Mg}{2} intensity between QS and CH (at the same magnetic field) could also be due to an instrumental effect. One candidate could be stray light. Also there could be another hitherto unknown process that would explain the difference in physical properties of the QS and CH chromosphere of the Sun. To test that there is no hidden bias we investigate the wing of the \ion{Mg}{2}~k line (i.e., integrated emission between 2832{--}2834~{\AA}). The wing of the \ion{Mg}{2}~k line represents the solar photosphere, where QS and CH intensities are very similar in QS and CH. The bottom panel of Fig.~\ref{fig:fig7} displays the same as that of the top and middle panels, but for the line wing. The overlap between CH and QS intensities is remarkable. If there are any differences between the two, then the CH actually appears slightly brighter then QS at similar magnetic flux. Therefore, we conclude that the difference between CH and QS seen in the k3 and k2v components of the line is not due to some hidden bias or instrumental effect.

\subsection{Effects of spectral resolution} 
Our analysis has demonstrated that there is a significant difference in narrow band \ion{Mg}{2}~k line core intensities between CH and QS. In this section, we follow two aims. Firstly, we investigate if the difference is revealed mainly by means of the unprecedented spectral resolution that is being provided by the IRIS spacecraft. Secondly, we want to check if CHs may influence the widely used \ion{Mg}{2} core-to-wing (c/w) ratio, also known as the \ion{Mg}{2} index \citep[see e.g.,][]{Heath1986}. For this purpose, we consider two scenarios. In the first scenario, we degrade the IRIS spectra to a resolution of 1~{\AA}, which is the spectral resolution of the SOLar-STellar Irradiance Comparison Experiment \citep[SOLSTICE;][]{McClintock2005} onboard SOLar Radiation and Climate Experiment \citep[SORCE;][]{Rottman2005} data used to determine the \ion{Mg}{2} index, and in the second scenario to 11~{\AA}, corresponding to the spectral resolution of the Solar Backscattered UltraViolet version~2 \citep[SBUV/2;][]{Fred1986,Cebula1992} on NOAA 16-18 data using a Gaussian smoothing function. After degrading the original IRIS spectra to 1~{\AA} and 11~{\AA}, the k2 and k3 features merge and cannot be well separated. The QS and CH intensities for each magnetic field bin are computed using the peak of each degraded spectrum for the 1~{\AA} scenario. However for the 11~{\AA} spectra, the intensities are estimated by taking the average at three different wavelength locations as suggested by \cite{Heath1986}. To improve the statistics, we used all the three data-sets listed in Table~\ref{table_obs}. For this purpose, we computed the intensities at the peak (1~{\AA} case) and at three different points (11~{\AA} case) and obtained an average of intensities within corresponding bins of $<$B$_{\rm LOS}$$>$/$\mu$. In order to increase the statistics, we have added all the three data sets described in Table~\ref{table_obs}.

\begin{figure*} 
\centering
\includegraphics[width=1.2\textwidth]{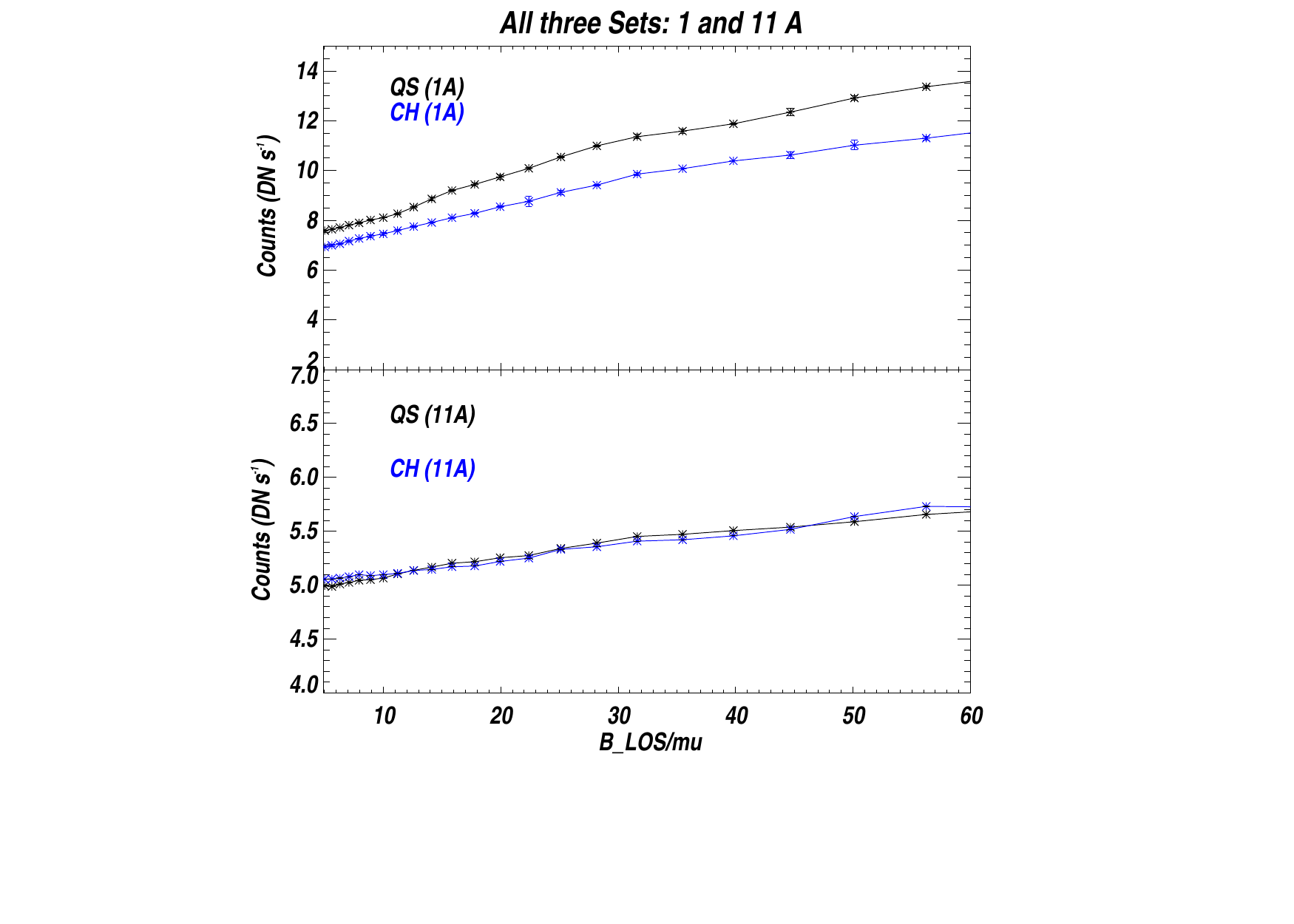}
\caption{Intensity vs. magnetic field based on low spectral resolution data. QS (black) and CH (blue) intensities as function of magnetic field estimated using 1~{\AA} spectra (top panel) and 11~{\AA} spectra (bottom panel).}
\label{fig:fig10}
\end{figure*}

We repeat the original analysis performed on \textsl{Data Set 2}, but now with reduced spectral resolution. Fig.~\ref{fig:fig10} shows the QS (black) and CH (blue) intensities averaged over magnetic field bins for 1~{\AA} (top panel) and 11~{\AA} resolution (bottom panel). These plots reveal that for 1~{\AA} resolution, the differences are still significant. However, for 11~{\AA} spectra, the difference in the intensities are minimal. While there is a significant intensity contrast between QS and CH at the original spectral resolution of IRIS, it is almost impossible to distinguish QS and CH at 11~{\AA} resolution. It is noted that changes in intensity with magnetic flux density in 11~{\AA} resolution spectra is small compared to that in 1~{\AA} resolution spectra. This is essentially due to the fact that while calculating the intensities for 1~{\AA} spectra the peak emission is taken, whereas for 11~{\AA} spectra average of three different locations were taken as per the definition by \cite{Heath1986}.

These results suggest that there is an influence of coronal holes on the \ion{Mg}{2} c/w ratio, at least if the more recently introduced version of the \ion{Mg}{2} index for spectral resolution of 1~{\AA} is considered \citep[see e.g.,][and references therein]{Snow2005}. Since the \ion{Mg}{2} core-to-wing ratio is often used as a proxy for plage areas \citep[see, e.g.,][]{Vie1999}, this intensity contrast between CH and QS must be taken into account in future work making use of the \ion{Mg}{2} index, in particular, if the study is based on spectra with 1~{\AA} resolution. 

\section{Discussion $\&$ Conclusions}\label{section:diss_conl}

IRIS provides high-resolution spectra of the \ion{Mg}{2} h and k resonance lines, which are important probes of the thermal and dynamic state of the upper solar chromosphere (\citealt{Leen2013}). The central absorption (k3) and short wavelength peak (k2v) of the \ion{Mg}{2}~k line form at different temperatures and altitudes in the solar chromosphere. In the current paper, we have studied the difference between the QS and CH intensities at different chromospheric layers using the intensities at the k2v and k3 wavelengths. Based on our analysis, we find that QS intensities in both k2v and k3 are significantly higher than those in CHs, when compared in regions of the same magnetic flux density ($<B_{LOS}>$/$\mu$). We also find that the difference in intensities is largest for k3, somewhat smaller for k2 and not significant in the wings of the Mg II k line. Consequently, this difference is only seen clearly if a high-resolution spectrum is used and that it decreases as the spectral resolution is lowered. Essentially, at 1~{\AA} resolution the difference between CH and QS is still visible, but for frequently used data for deriving the \ion{Mg}{2} index at 11~{\AA}, it is not. 

Generally, it is reported that the QS and CH start to show intensity contrast only if the temperature exceeds 6$\times$ 10$^{5}$~K (\citealt{Wilhelm2000, Stucki2000, Stucki2002, Cranmer2009}). Therefore, it is often difficult to isolate CH regions from QS without taking recourse to coronal images. The main spectral lines formed at lower temperatures that clearly show a contrast between CH and QS are lines of \ion{He}{1} and \ion{He}{2}, e.g., the \ion{He}{1}~10830~{\AA} absorption line (\citealt{Harvey1975, Harvay2002}). However, the \ion{He}{1}~10830~{\AA} line is affected by coronal radiation. Also, the \ion{C}{4} line at 1548~{\AA} has been reported to show a lower intensity in a polar coronal hole \citep[][]{Peter_1999b}.

The question arises why the intensities in QS and CH are so different in the \ion{Mg}{2}~k line core. The difference between QS and CH in coronal radiation was explained by \cite{Wie2004} by invoking loop statistics and Rosner-Tucker-Vaiana \citep[RTV;][]{rtv} scaling. In CHs the longer loops, which reach higher altitudes and get hotter are missing. This explains why CHs are so dark compared to the QS at coronal temperatures. Shorter, less hot loops are almost as common in CHs as in the QS, which is why no (or only a small) intensity contrast between CH and QS is expected for the chromosphere. However, we note that this explanation is not directly applicable in the chromosphere due to the fact that the RTV scaling used in that paper does not apply. Further work is required to explain the difference in \ion{Mg}{2} core brightness in CH and QS.

The core-to-wing ratio of the \ion{Mg}{2} lines, known as the \ion{Mg}{2}-index \citep[see, e.g.][]{Heath1986} are taken as a proxy for solar activity in modelling the total and spectral solar irradiance (TSI and SSI), in particular in plage areas. In the past, the index has been derived using 11~{\AA} and 1~{\AA} resolution spectra that were obtained considering Sun-as-a-star data \cite[see e.g.,][]{Heath1986,DeToma1997,Vie2001,Snow2005,Snow2014}. Our findings suggest that the \ion{Mg}{2} index for CHs should be lower than in QS, in particular for 1{\AA} resolution data. This fact has not been accounted for in the modelling of the total solar irradiance (TSI) or SSI using the Mg index so far. In the future, the Solar Ultraviolet Imaging Telescope \cite[SUIT;][]{GhoCKT_2016, Trip_suit} on board the Aditya-L1 mission of the Indian Space Research Organisation (ISRO) shall provide full disk images of the Sun in 11 different passbands in the 2000{--}4000~{\AA} wavelength range, including two filters centered at the \ion{Mg}{2}~h\&k lines and two at their wings. The images in the \ion{Mg}{2} h \& k lines and their wings will be obtained with a spectral resolution of 4~{\AA}. These observations will for the first time, provide opportunities to measure the spatially resolved \ion{Mg}{2} indices for the whole Sun, which has not been possible so far. This in turn will provide improved empirical models of the TSI as well as SSI.

\acknowledgements We thank the anonymous referee for the constructive comments. This research is supported by the Max-Planck Partner Group of MPS at IUCAA that is funded by MPG and DST (IGSTC). IRIS is a NASA small explorer mission developed and operated by LMSAL with mission operations executed at NASA Ames Research center and major contributions to downlink communications funded by ESA and the Norwegian Space Centre. AIA and HMI data are courtesy of NASA/SDO and the AIA and HMI science teams.This work was partly supported by the BK21 plus program through the National Research Foundation (NRF) funded by the Ministry of Education of Korea. PK is currently funded through the National Science Centre, Poland, (NCN) grant no. 2014/15/B/ST9/00106.

\begin{table*}
\centering
\caption{Summary of all the three sets of observations of quiet Sun (QS) and coronal holes (CH). $\tau$ is the exposure time, $\mu$ is the cosine of the heliocentric angle and B$_{LOS}$ is the average LOS magnetic field. \label{table_obs}} 
\begin{tabular}{|c|c|c|c|c|c|c|c|c|c|c|}
\hline
 				& \multicolumn{2}{c|}{\footnotesize{Date(Time)}} 			& \multicolumn{2}{c|}{\footnotesize{Field of View}} & \multicolumn{2}{c|}{$\tau$ [s]} & \multicolumn{2}{c|}{$\mu$} & \multicolumn{2}{c|}{B$_{LOS}$/$\mu$ [G]}\\
 				\cline{2-11}
Data		&	QS		&	CH		&	QS		&	CH		&	QS		&	CH		&	QS		&	CH & QS		&	CH\\
\hline
Set 1       & \multicolumn{2}{c|}{10.11.15(13:30:04)}        	& 		\multicolumn{2}{c|}{129\arcsec $\times$ 170\arcsec}			& \multicolumn{2}{c|}{15}  & 0.81 & 0.88 & 10.06 & 10.64 \\
				\hline
Set 2    & 04.01.16   	& 05.01.16    & 	126\arcsec $\times$ 130\arcsec & 129\arcsec $\times$173\arcsec & 15  & 15	 & 0.90  & 0.90 & 10.18 & 10.41\\
			 &  (04:54:38)		& (14:50:33)	&			&&&&&&&																																																							\\
\hline
Set 3    & 18.12.14 	& 29.11.14  & 34\arcsec $\times$	181.0\arcsec & 126\arcsec$\times$129.0\arcsec & 30 & 30  & 0.88 & 0.81 & 19.44 & 16.98 \\
			 & (05:47:42)	& (23:02:45) &&&&&&&&\\
\hline
\end{tabular}
\end{table*}
\begin{table*}
\centering
\caption{Number of \ion{Mg}{2} profiles in each magnetic field bin. \label{table_profile}} 
\begin{tabular}{|c|c|c|}
\hline
          & \multicolumn{2}{c|}{\footnotesize{\ion{Mg}{2} k (No. of Profiles)}} \\
                               \cline{2-3}
B$\_$LOS Bins	&	QS		&	CH\\
\hline
3.98-4.46   & 1658 & 772\\
\hline
4.46-5.01   & 1857 & 971\\
\hline
5.01-5.62   & 2030 & 1013\\
\hline
5.62-6.30   & 2072 & 1032\\
\hline
6.30-7.07   & 2059 & 1094\\
\hline
7.07-7.94   & 2064 & 1009\\
\hline
7.94-8.91   & 1740 & 894\\
\hline
8.91-10.00   & 1485 & 762\\
\hline
10.00-11.22   & 1125 & 680\\
\hline
11.22-12.58   & 923 & 484\\
\hline
12.58-14.12   & 716 & 369\\
\hline
14.12-15.84   & 567 & 301\\
\hline
15.84-17.78   & 411 & 254\\
\hline
17.78-19.95   & 310 & 156\\
\hline
19.95-22.38   & 262 & 111\\
\hline
22.38-25.12   & 208 & 76\\
\hline
25.12-28.18   & 178 & 61\\
\hline
28.18-31.62   & 157 & 39\\
\hline
31.62-35.48   & 111 & 36\\
\hline
35.48-39.81   & 113 & 33\\
\hline
39.81-44.66   & 80 & 47\\
\hline
44.66-50.12   & 71 & 33\\
\hline
50.12-56.32   & 60 & 30\\
\hline
56.32-63.09   & 53 & 33\\
\hline
\end{tabular}
\end{table*}

\clearpage

\clearpage

\bibliographystyle{apj} 
\bibliography{reference} 


\newpage

\appendix

\section{Data Set 3: The observations on 29 Nov and 18 Dec 2014}\label{appendix:A}
\begin{figure*} 
\centering
\includegraphics[width=0.7\textwidth]{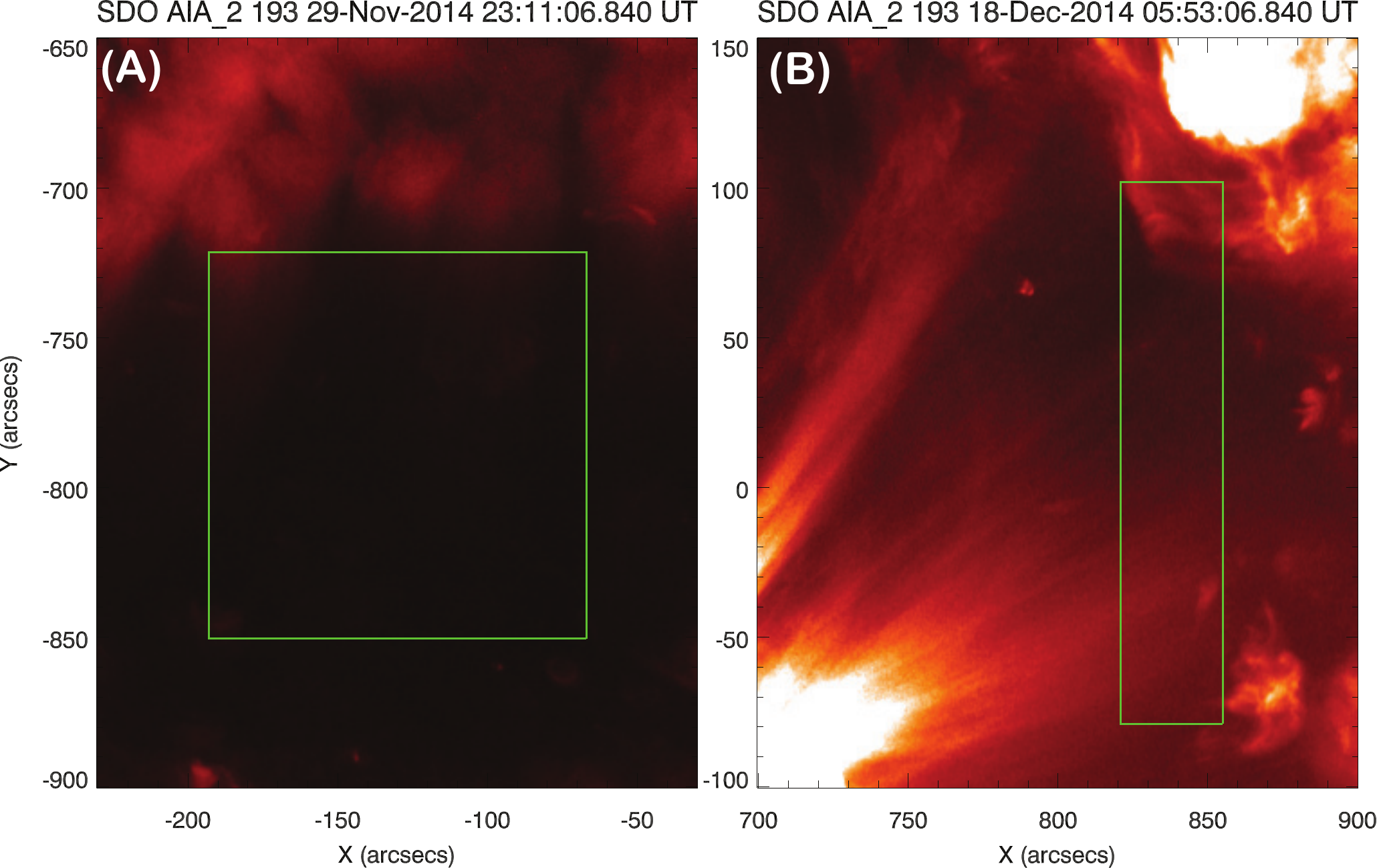}
\caption{Fields-of-view for data set 3 as seen in the 193~{\AA} channel of AIA (similar to Fig~\ref{fig:fig4}).}\label{fig:fig11}
\end{figure*}

\begin{figure*} 
\centering
\includegraphics[width=1.2\textwidth]{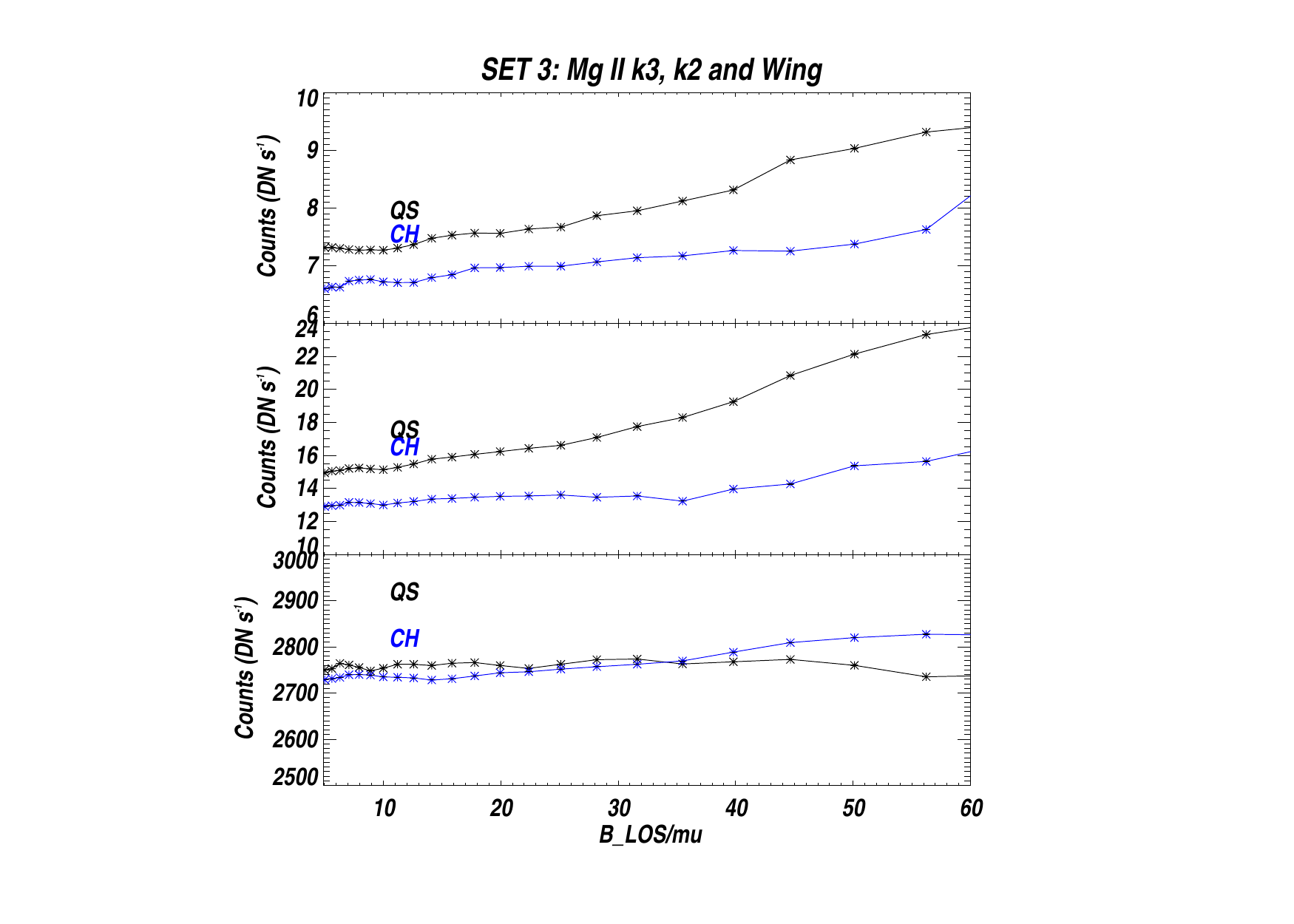}
\caption{Same as Fig.~\ref{fig:fig7} but for data set 3.}
\label{fig:fig16}
\end{figure*}

Here we present the results obtained from the date set 3 as listed Table~\ref{table_obs}. The QS observation is a 18 step coarse raster with an exposure time of 30~s, while the CH observation is a 64 step coarse raster with an exposure time of 15~s. The $\mu$ range covered in the CH study is rather large (0.449 to 0.659) as compared to that of the QS (0.471 to 0.539). Therefore, the analysis was performed to only a selected part of the QS study where the $\mu$ values overlapped. The fields-of-view for this case are shown in Fig.~\ref{fig:fig11} and the resulting chromospheric emission vs. magnetic field is displayed in Fig.~\ref{fig:fig16}

\end{document}